\documentclass[12pt]{article}
\usepackage{a4wide}

\usepackage[dvipdfmx,hiresbb]{graphicx} 
\usepackage{mediabb}                    
\usepackage[
  top=30mm,
  bottom=20mm,
  left=20mm,
  right=20mm
]{geometry}

\usepackage{amsmath,amssymb}
\usepackage{color}
\usepackage[bookmarks,bookmarksnumbered]{hyperref}
\usepackage{cellspace}
\usepackage{tikz}
\usepackage{mathtools}
\usepackage{cite}
\usetikzlibrary{fadings}
\usetikzlibrary{patterns}
\usetikzlibrary{shadows.blur}
\usetikzlibrary{shapes}
\usepackage{ascmac}
\usepackage{mathrsfs}
\usepackage{float}
\usepackage{adjustbox}

\newcommand{\aln}[1]{\begin{align}#1\end{align}}
\newcommand{\nn}{\nonumber\\}

\begin{document}
\title{\vspace{-3cm}
\vbox{
\baselineskip 14pt
\hfill \hbox{\normalsize 
}}
\vskip 1cm 
\bf \Large Variational Method for Interacting Surfaces with Higher-Form Global Symmetries 
\vskip 0.5cm
}
\author{
Kiyoharu Kawana\thanks{E-mail: \tt kkiyoharu@kias.re.kr}
\bigskip\\
\normalsize
\it School of Physics, Korean Institute for Advanced Study, Seoul 02455, Korea
\smallskip
}
\date{\today}

\maketitle   
 
\begin{abstract} 
We develop a variational method for interacting surface systems with higher-form global symmetries.   
As a natural extension of the conventional second-quantized Hamiltonian of  interacting bosons, we explicitly construct a second-quantized Hamiltonian formulated in terms of a closed surface operator $\hat{\phi}[C_p^{}]$ charged under a $p$-form global symmetry. 
Applying the variational principle, we derive a functional Schr\"{o}dinger equation analogous to the Gross-Pitaevskii equation in conventional bosonic systems.   
In the absence of external forces, the variational equation admits a uniform solution that is uniquely determined by a microscopic interaction potential $U(\psi^*\psi)$ and the chemical potential. 
This uniform solution describes a uniform gas of bosonic surfaces.  
Using the obtained energy functional, we show that low-energy fluctuations contain a gapless $p$-form field $A_p^{}$ when the $p$-form global symmetry is $\mathrm{U}(1)$, whereas the $p$-form field becomes massive for discrete symmetries, whose low-energy limit is described by a $\mathrm{BF}$-type topological field theory. 
As a consequence, the system exhibits abelian topological order with anyonic surface excitations.   
In the presence of external forces, however, solving the functional equation in full generality remains challenging.  
We argue, however, that the problem reduces to solving the conventional Gross-Pitaevskii  equation  when external forces act separately on the center-of-mass and relative motions.       
In addition, we present analytic solutions for topological defects as analogs of vortex and domain-wall solutions in conventional bosonic systems.         
Finally, as a concrete microscopic model, we study a $\mathbb{Z}_N^{}$ lattice gauge theory and apply our variational method to this system. 

\end{abstract} 

\setcounter{page}{1}

\newpage

\tableofcontents   

\section{Introduction}\label{Sec:intro}

One of the central themes in modern physics is to understand and classify quantum phases of matter from the perspective of generalized symmetries~\cite{Gaiotto:2014kfa,Kapustin:2005py,Pantev:2005zs,Nussinov:2009zz,Banks:2010zn,Kapustin:2013uxa,Aharony:2013hda,Kapustin:2014gua,Gaiotto:2017yup,
McGreevy:2022oyu,
Brennan:2023mmt,Bhardwaj:2023kri,Luo:2023ive,Gomes:2023ahz,Shao:2023gho,bhardwaj2023lecturesgeneralizedsymmetries
}. 
The recognition of generalized symmetries has significantly broadened our understanding of quantum phases of matter and motivated the  exploration of new paradigms beyond conventional Landau theories~\cite{Iqbal:2021rkn,Hidaka:2023gwh,Pace:2023ccj,Liu:2024znj,Kawana:2024fsn,Kawana:2024qmz,Kawana:2025vvf,Kawana:2025vbi}.       

In this paper, we investigate a variational method for interacting bosonic surfaces with higher-form global symmetries. 
In the case of conventional $0$-form global symmetries, the variational method is well established and has been applied to a variety of interacting bosonic systems in the analysis of ground states and low-energy excitations~\cite{RevModPhys.71.463,RevModPhys.73.307,Pethick_Smith_2008,Griffin_Nikuni_Zaremba_2009,Yukalov:2011qj,Proukakis}.  
The primary objective of this work is to extend this conventional variational method to nonrelativistic bosonic systems possessing higher-form global symmetries. 

In Section~\ref{Variational principle}, we begin by reviewing the standard variational method in interacting surface systems with a $\mathrm{U}(1)$ $0$-form global symmetry. 
Readers familiar with this subject may skip this part. 
We then generalize the variational method to interacting surface systems with $\mathrm{U}(1)$ higher-form global symmetries.  
We explicitly construct a second-quantized Hamiltonian of a $p$-dimensional closed surface operator $\hat{\phi}[C_p^{}]$ on a $(D-1)$-dimensional spatial hypercubic lattice.    
By evaluating the expectation value of the Hamiltonian in a coherent state built from the surface operator, we obtain a free-energy functional of the variational wave function $\psi_G^{}[C_p^{}]$. 
The minimization of this functional leads to a functional Schr\"{o}dinger equation analogous to the Gross-Pitaevskii equation in conventional bosonic systems~\cite{RevModPhys.71.463,RevModPhys.73.307}.  
Throughout the paper, we refer to this functional equation as the generalized Gross-Pitaevskii equation.   
In addition, we discuss several distinctive features that are unique in systems of interacting surfaces in contrast to conventional bosonic particles.

In Section~\ref{mean field analysis}, we perform a mean-field analysis based on the obtained free-energy functional and the generalized Gross-Pitaevskii equation. 
In the absence of external forces, the system describes a uniform gas of interacting surfaces, whose ground state is determined by a  microscopic interaction potential $U(\hat{\phi}^\dagger \hat{\phi})$ in close analogy with conventional bosonic systems.   
We discuss the condensation criterion in detail and argue that (non-)condensed phase is characterized by the longe-range order (the area law) of the expectation value of the surface operator $\langle \hat{\phi}[C_p^{}]\rangle$. 
In the condensed phase, i.e, the broken phase of $\mathrm{U}(1)$ $p$-form symmetry, we explicitly show that low-energy excitations contain a gapless higher-form field $A_p^{}$ as an analog of sound-wave modes in conventional bosonic systems. 

In the presence of external forces, on the other hand, solving the generalized Gross-Pitaevskii equation in full generality remains a challenging task.     
Here, we point out that substantial simplification of the problem can occur when the external force $V[C_p^{}]$ takes a separable form as $V[C_p^{}]=V_{\rm CM}^{}(x)+V_R^{}[\{Y\}]$, where $x$ denotes the center-of-mass position and $\{Y\}=\{Y^k(\xi)\}_{k=1}^{D-1}$ represent the relative coordinates of the surface. 
In this case, by employing the method of separation of variables, the free-energy functional can be reduced to an effective free-energy functional of  interacting bosons, where the effects of internal degrees of freedom are entirely absorbed into renormalized effective couplings.     
After this general argument, we study a spherically symmetric  harmonic potential acting on the center-of-mass position and find that the thermodynamic properties of the ground state are completely identical to those of  conventional trapped bosons within the Thomas-Fermi approximation.    
We can also study the spontaneous breaking of discrete higher-form global symmetries within the present framework. 
In the second quantization picture, a discrete symmetry can be realized by adding potential terms such as $\hat{\phi}[C_p^{}]^N+{\rm h.c.}$ to the Hamiltonian.  
The inclusion of such terms lifts the infinite ground-state degeneracy and leaves a discrete set of ground states. 
As a result, the system exhibits abelian topological order in the condensed phase.     
We explicitly derive the low-energy effective theory and find that it is described by a $\mathrm{BF}$-type topological field theory.  
Besides, we present a detailed discussion of ground-state degeneracy and anyonic surface excitations in the present framework.  
In the last part of Section~\ref{mean field analysis}, we also provide analytic solutions of topological defects, which can be viewed as natural generalizations of vortex and domain-wall solutions in conventional bosonic systems.   

In Section~\ref{sec:model}, we study a $\mathbb{Z}_N^{}$ lattice gauge theory as a concrete microscopic model and apply our variational method.  
In the first subsection, we describe the model and higher-form (gauge) symmetries in detail and emphasize its relation to the $\mathbb{Z}_N^{}$ toric code~\cite{Kitaev:1997wr,Bullock:2006bv,Zou:2016dck,Watanabe:2021wwt} as well. 
Then, the ground states are explicitly constructed in the strong and weak gauge coupling limits respectively.  
While the ground state in the strong gauge coupling limit corresponds to a non-condensed phase of surfaces, the ground state in the weak gauge-coupling limit represents a condensation of surfaces analogous to the string-net condensation in the $\mathbb{Z}_2^{}$ toric code in $1+2$ dimension~\cite{Yoshida:2013sqa,Wen:2003yv}. 
We then apply our variational method to this system and find that the resulting free-energy functional takes a form analogous to that we constructed in the previous sections, up to a minor difference in the kinetic term. 
This allows us to apply the mean-field analysis developed in Section~\ref{mean field analysis} and to investigate a variety of low-energy properties of the gauge theory, including the ground-state structure, low-energy effective theory, anyonic excitations, topological defects, and so on.  
In Section~\ref{sec:summary}, we present summary and discussion.  

Throughout this paper, we denote a $(D-1)$-dimensional spatial manifold by $\Sigma_{D-1}^{}$ and its metric by $g_{ij}^{}(x)$. 
The corresponding spacetime manifold is denoted by $\Sigma_D^{}=\mathbb{R}\times \Sigma_{D-1}^{}$. 

\section{Variational method}\label{Variational principle}
We develop a variational method for nonrelativistic bosonic systems with higher-form global symmetries.   
To distinguish different higher-form symmetries, we denote a $p$-form global symmetry generally as $G^{[p]}$, which is an Abelian group for $p\geq 1$.   
For $p=0$, $G^{[0]}$ can be non-Abelian. 
%

\subsection{Conventional variational method}
Let us recap the variational method in conventional bosonic systems with an abelian $0$-form global symmetry $G^{[0]}$.   
The existence of the symmetry $G^{[0]}$ implies that there exists an $(D-1)$-dimensional symmetry (or topological) operator $\hat{U}_\theta^{}[\Sigma_{D-1}^{}]$ that commutes with a Hamiltonian $\hat{H}$,   
\aln{
[\hat{U}_\theta^{}[\Sigma_{D-1}^{}],\hat{H}]=0~,
\label{commutative equation}
}
where $\theta $ is a transformation parameter. 
When $G^{[0]}=\mathrm{U}(1)^{[0]}$, the symmetry operator is associated with a $\mathrm{U}(1)$ conserved charge:
\aln{\hat{U}_\theta^{}[\Sigma_{D-1}^{}]=e^{i\theta \hat{Q}[\Sigma_{D-1}^{}]}~,\quad \hat{Q}[\Sigma_{D-1}^{}]=\int_{\Sigma_{D-1}^{}}\star \hat{J}_1^{}~, 
}
where $\hat{J}_1^{}$ is a conserved $1$-form current, $d\star \hat{J}_1^{}=0$.   
Typically, such a theory 
contains a charged local operator $\hat{\phi}(x)$ satisfying  
\aln{
\hat{U}_\theta^{}[\Sigma_{D-1}^{}]\hat{\phi}(x)\hat{U}_\theta^{-1}[\Sigma_{D-1}^{}]=e^{iq\theta}\hat{\phi}(x)~,
}
where $q$ is a charge (i.e. representation) under $G^{[0]}$. 

Suppose that we are given a Hamiltonian $\hat{H}$ that (i) is constructed from $\hat{\phi}(x)$ and (ii) satisfies Eq.~(\ref{commutative equation}).  
The primary objective is to identify a ground state and to understand its various  phases.    
For this purpose, we introduce the {\it trivial state} $|0\rangle$ via 
\aln{
\hat{\phi}(x)|0\rangle=0\quad \text{for } ^\forall x\in \Sigma_{D-1}^{}~,
}
and define a variational state by 
\aln{
|\psi_G^{}\rangle\coloneq \sqrt{{\cal N}}\exp\left(\int_{\Sigma_{D-1}^{}} d^{D-1}x\sqrt{g(x)}\psi_G^{}(x)\hat{\phi}^\dagger (x)\right)|0\rangle~,
\label{particle condensed state}
}
where $\psi_G^{}(x)$ is a variational wave function, $g(x)$ is the  determinant of the spatial metric $g_{ij}^{}(x)$, and $\sqrt{{\cal N}}$ is a normalization factor.~(Here, the subscript $G$ means ground state.)
When $\Sigma_{D-1}^{}$ is a discretized space such as a hypercubic lattice, this state can be also written as  
\aln{
|\psi_G^{}\rangle=\underset{x\in \Sigma_{D-1}^{}}{\otimes} |\psi_G^{}(x)\rangle~,
}
that is, it is a product state of local coherent states $|\psi_G^{}(x)\rangle$. 

Taking the expectation value of $\hat{H}$ and $\hat{Q}$, we obtain the free-energy functional 
\aln{F[\psi_G^{}]\coloneq \langle \psi_G^{}|\hat{H}-\mu(\hat{Q}-Q)|\psi_G^{}\rangle~,
} 
where $\mu$ is the chemical potential. 
Then, the minimization of the free energy 
\aln{
\frac{\delta F[\psi_G^{}]}{\delta \psi_G^*(x)}=0~
}
leads to the variational equation along with the charge condition 
\aln{\frac{\partial F[\psi_G^{}]}{\partial \mu}=0
\quad \leftrightarrow \quad Q=\langle \psi_G^{}|\hat{Q}|\psi_G^{}\rangle~.
}
As an example, let us consider the following Hamiltonian:  
\aln{
\hat{H}=\int_{\Sigma_{D-1}^{}} d^{D-1}x\sqrt{g(x)}\left[\hat{\phi}^\dagger(x)\left(-\frac{1}{2m}\Box+V(x)\right)\hat{\phi}(x)+U(\phi^\dagger \phi)
\right]~,
\label{particle Hamiltonian}
}
where 
\aln{\Box=\frac{1}{\sqrt{g(x)}}\partial_i^{}(\sqrt{g(x)}g^{ij}(x)\partial_j^{}) 
}
is the Laplace operator, and the field operator $\hat{\phi}(x)$ satisfies the commutation relation
\aln{\label{bosonic commutation}
[\hat{\phi}(x),\hat{\phi}^\dagger(y)]=\frac{1}{\sqrt{g(x)}}\delta^{(D-1)}(x-y)~.
}
Here, $V(x)$ is an external potential which is independent of the field.  
It is straightforward to check $[\hat{U}_\theta^{}[\Sigma_{D-1}^{}],\hat{H}]=0$ and the $\mathrm{U}(1)^{[0]}$ transformation 
\aln{
\hat{U}_\theta^{}[\Sigma_{D-1}^{}]\hat{\phi}(x)\hat{U}_\theta^{-1}[\Sigma_{D-1}^{}]=e^{i\theta}\hat{\phi}(x)~
}
with 
\aln{
\hat{U}_\theta^{}[\Sigma_{D-1}^{}]=e^{-i\theta \hat{Q}[\Sigma_{D-1}^{}]}~,\quad \hat{Q}[\Sigma_{D-1}^{}]\coloneq \int_{\Sigma_{D-1}^{}}d^{D-1}x\sqrt{g(x)}~\hat{\phi}^\dagger(x)\hat{\phi}(x)~. 
\label{0-form charge}
}
In this case, the free-energy functional is 
\aln{
F[\psi_G^{}]&=E[\psi_G^{}]-\mu (Q[\psi_G^{}]-Q)
\nn
&=\int_{\Sigma_{D-1}^{}} d^{D-1}x\sqrt{g(x)}\left[\psi_G^{}(x)^*\left(-\frac{1}{2m}\Box+V(x)-\mu\right)\psi_G^{}(x)+U(\psi_G^* \psi_G^{})\right]+\mu Q~.
\label{free energy functional}
}  
Taking the variation with respect to $\psi_G^{}(x)^*$, we obtain the variational equation
\aln{
0&=\left(-\frac{1}{2m}\Box+V(x)-\mu\right)\psi_G^{}(x)+\frac{\delta U(\psi_G^* \psi_G^{})}{\delta \psi_G^{}(x)^*}~,\quad Q=\int_{\Sigma_{D-1}^{}} d^{D-1}x |\psi_G^{}(x)|^2~,
\label{GP equation}
} 
which is known as the Gross-Pitaevskii equation in condensed matter physics~\cite{RevModPhys.71.463,RevModPhys.73.307,Pethick_Smith_2008}. 

Several comments are in order. 
First, the field operator can be generally expanded as 
\aln{
\hat{\phi}(x)=\sum_{i}\hat{a}_i^{}\psi_i^{}(x)~,
\label{function expansion}
}
by an orthonormal set of functions $\{\psi_0^{}=\psi_G^{}/\sqrt{Q},\psi_k^{}\}$ satisfying 
\aln{
\int_{\Sigma_{D-1}^{}} d^{D-1}x\sqrt{g(x)}\psi_i^{}(x)^*\psi_j^{}(x)=\delta_{ij}^{}~,
}
and 
\aln{[\hat{a}_i^{},\hat{a}_j^\dagger]=\delta_{ij}^{}~,\quad \text{others}=0~.
\label{harmonic commutation}
}
Then, Eq.~(\ref{particle condensed state}) is equivalently written as
\aln{
|\psi_G^{}\rangle=\sqrt{{\cal N}}e^{\sqrt{Q}\hat{a}_0^\dagger}|0\rangle~, 
}
which implies that the ground state is a coherent state of $\hat{a}_0^{}$ and $\hat{a}_0^\dagger$. 

Second, we can also derive the action which leads to the Hamiltonian~(\ref{particle Hamiltonian}) and the commutation relation~(\ref{bosonic commutation}) as  
\aln{
\label{particle action}
S[\phi]=\int_{\Sigma_{D}^{}}\star 1~\left[i(\phi(t,x)^*\dot{\phi}(t,x)-\phi(t,x)\dot{\phi}(t,x)^*)-\phi(t,x)^*\left(-\frac{1}{2m}\Box+V(x)\right)\phi(t,x)-U(\phi^*\phi)
\right]~,
}
whose equation motion $\frac{\delta S[\phi]}{\delta \phi^*(x)}=0$ describes a dynamics of the condensate, i.e., the time-dependent Gross-Pitaevskii equation.

\subsection{Generalization to surface systems} 

\begin{figure}
    \centering
    \includegraphics[scale=0.7]{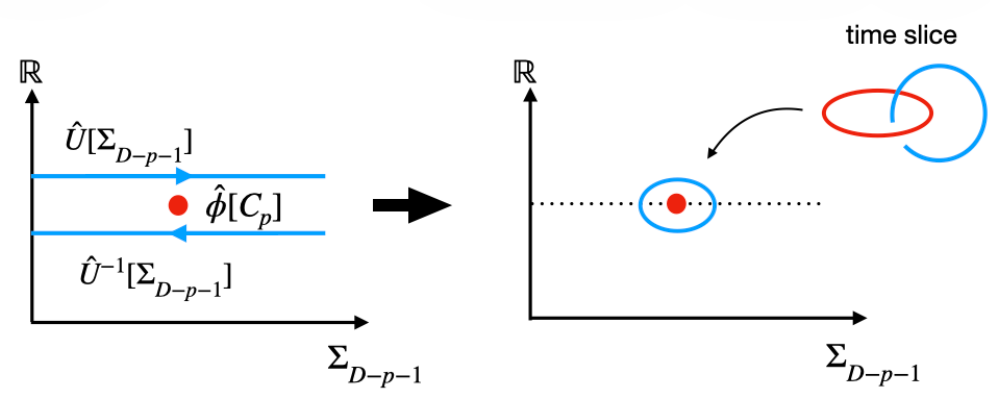}
    \caption{
  Two surface operators that lead to the commutation relation~(\ref{p-form symmetry}). 
    The arrows indicate the orientation of the subspaces.
    }
    \label{fig:link}
\end{figure}

Now, let us consider many-body bosonic systems with higher-form global symmetries.   
A $p$-form global symmetry $G^{[p]}$ implies that there exists a $(D-p-1)$-dimensional symmetry (or topological) operator $\hat{U}_\theta^{}[\Sigma_{D-p-1}^{}]$ that commutes with a Hamiltonian, 
\aln{
[\hat{U}_\theta^{}[\Sigma_{D-p-1}^{}],\hat{H}]=0~,
}
where $\Sigma_{D-p-1}^{}$ is a $(D-p-1)$-dimensional closed subspace in $\Sigma_{D-1}^{}$ and $\theta $ is a transformation parameter.   
For $G^{[p]}=\mathrm{U}(1)^{[p]}$, the symmetry operator is associated with the conserved charge as 
\aln{
\hat{U}_\theta^{}[\Sigma_{D-p-1}^{}]=e^{-i\theta \hat{Q}[\Sigma_{D-p-1}^{}]}~,\quad \hat{Q}_p^{}[\Sigma_{D-p-1}^{}]=\int_{\Sigma_{D-p-1}^{}}\star J_{p+1}^{}~,
}
where $J_{p+1}^{}$ is a $(p+1)$-form conserved current, namely, it satisfies $d\star \hat{J}_{p+1}^{}=0$. 
Analogous to the $0$-form case, such a system typically contains a charged  
$p$-dimensional surface operator $\hat{\phi}[C_p^{}]$ that obeys an equal-time operator relation 
\aln{
\hat{U}_\theta^{}[\Sigma_{D-p-1}^{}]\hat{\phi}[C_p^{}]\hat{U}_\theta^{-1}[\Sigma_{D-p-1}^{}]=e^{iq\theta \mathrm{I}[C_p^{},\Sigma_{D-p-1}^{}]}\hat{\phi}[C_p^{}]~,
\label{p-form symmetry}
}  
where $C_p^{}$ is an embedded surface of $S^p$ in $\Sigma_{D-1}^{}$,  and $\mathrm{I}[C_p^{},\Sigma_{D-p-1}^{}]$ is the intersection number between them: 
\aln{
\mathrm{I}[C_p^{},\Sigma_{D-p-1}^{}]\coloneq \lim_{\epsilon\rightarrow 0}\int_{\Sigma_D^{}}\delta_{D-p}^{}(C_{p}^{})\wedge \delta_{p}^{}(M_{D-p}^{\varepsilon})~,
}
where $M_{D-p}^{}$ is the $(D-p)$-dimensional spacetime region enclosed by $C_{D-p-1}^{\varepsilon}\coloneq \Sigma_{D-p-1}^{}(t+\epsilon)\cup \overline{\Sigma}_{D-p-1}^{}(t-\epsilon)$, and $\delta_{D-q}^{}({\cal M}_q^{})$ is the Poincare dual form of a $q$-dimensional manifold ${\cal M}_q^{}$ in general.  
See Fig.~\ref{fig:link} for example. 

%
The trivial state $|0\rangle$ is similarly defined by 
\aln{\hat{\phi}[C_p^{}]|0\rangle=0\quad \text{for } ^\forall C_p^{}~,
}
and a variational state is introduced by 
\aln{
|\psi_G^{}\rangle=\sqrt{{\cal N}}:\exp\left({\int}{\cal D}Xw[C_p^{}]~\hat{\phi}^\dagger [C_p^{}]\psi_G^{}[C_p^{}]
\right):|0\rangle~,
\label{surface condensed state}
}  
where $\psi_G^{}[C_p^{}]$ is a variational wave function, $\int {\cal D}X$ is the path-integral over all embeddings $C_p^{}=\{X^i(\xi)\}_{i=1}^{D-1}:S^p\rightarrow \Sigma_{D-1}^{}$, $w[C_p^{}]$ is a weight functional, and $:\cdots:$ denotes a prescription for properly accounting for identical states when expanding the exponential in general. 
See the next subsection about the weight functional $w[C_p^{}]$.  

The free-energy functional is now given by 
\aln{
F[\psi_G^{};\Sigma_{D-p-1}^{}]=\langle \psi_G^{}|\hat{H}-\mu_p^{}(\hat{Q}_p^{}[\Sigma_{D-p-1}^{}]-Q_p^{})|\psi_G^{}\rangle~,
\label{higher-form energy functional}
}
where $\mu_p^{}$ is the chemical potential for the $p$-form global charge. 
Taking the variations with respect to $\psi_G^{}[C_p^{}]^*$ and $\mu_p^{}$, we have 
\aln{
\frac{\delta F[\psi_G^{};\Sigma_{D-p-1}^{}]}{\delta \psi_G^{}[C_p^{}]^*}=0~,\quad \langle \psi_G^{}|\hat{Q}_p^{}[\Sigma_{D-p-1}^{}]|\psi_G^{}\rangle=Q_p^{}~.
}
Here, a couple of comments are in order. 
First, a concrete microscopic system can also possess additional higher-form global symmetries $\mathrm{U}[1]^{[q]}$ with $q\leq p$.       
In such a case, we can also add another Lagrange term $\mu_q^{}(\hat{Q}_q^{}-Q_q^{})$ to the above free energy.  

Second, when the spatial manifold $\Sigma_{D-1}^{}$ is topologically trivial, i.e., $\Sigma_{D-1}^{}=\mathbb{R}^{D-1}$, the $p$-form global charge $Q_p^{}$ cannot take a nonzero value since $\mathrm{I}[C_p^{},\Sigma_{D-p-1}^{}]=0$ for all $C_p^{}$ and $\Sigma_{D-p-1}^{}$. 
In this case, Eq.~(\ref{higher-form energy functional}) becomes independent of $\Sigma_{D-p-1}^{}$, and condensation of the system is governed by the coupling constants in $\hat{H}$.   
%
%
On the other hand, when the topology of $\Sigma_{D-1}^{}$ is nontrivial so  that it contains topologically nontrivial subspaces along which $C_p^{}$ and/or $\Sigma_{D-p-1}^{}$ can wind, the ground state can carry a nonzero $p$-form global charge $Q_p^{}\neq 0$.  

Let us present a concrete example analogous to Eq.~(\ref{free energy functional}) next.

\subsection{A lattice model of bosonic surfaces}

\begin{figure}
    \centering
    \includegraphics[scale=0.6]{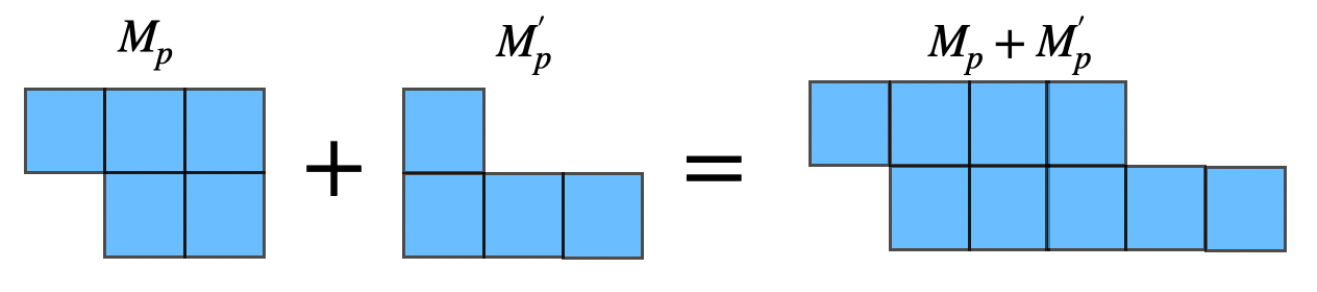}
    \caption{
 A fusion of two surfaces. 
    }
    \label{fig:fusion}
\end{figure}

In order to introduce a concrete model, let us consider a hypercubic spatial lattice $\Sigma_{D-1}^{}=\Lambda_{D-1}^{}$ with a lattice spacing $a$ and the periodic boundary condition.  
On this discretized space, a $p$-dimensional closed surface $C_p^{}$ is constructed by gluing multiple $p$-dimensional minimum hypercubes $\sigma_{p}^{}$ whose topology is $p$-dimensional ball.  
In particular, when $C_p^{}$ is constructed by $N$ hypercubes, its volume is
\aln{
\mathrm{Vol}[C_p^{}]=a^{p}N~,
} 
and its continuum limit is given by $a\rightarrow 0~,N\rightarrow \infty$ with $\mathrm{Vol}[C_p^{}]=\text{fixed}$.  
On each $\sigma_{p}^{}$, we assign a Hilbert space 
 \aln{
 {\cal H}_{\mathbb{C}}\coloneq \text{span}\{|\phi(\sigma_p^{})\rangle~,\quad \phi(\sigma_p^{})\in \mathbb{C}\} 
 }
 where $|\phi(\sigma_p^{})\rangle$ is the eigenstate of a local operator $\hat{\phi}(\sigma_p^{})$, i.e., $\hat{\phi}(\sigma_p^{})|\phi(\sigma_p^{})\rangle=\phi(\sigma_p^{})|\phi(\sigma_p^{})\rangle$. 
Besides, we assume 
a commutation relation
\aln{
[\hat{\phi}(\sigma_p^{}),\hat{\phi}^\dagger(\sigma_p^{'})]=w[\sigma_p^{}]^{-1}\delta_{\sigma_p^{},\sigma_p^{'}}^{}=\begin{cases} w[\sigma_p^{}]^{-1} & \text{for } \sigma_p^{}=\sigma_p^{'}
\\
0 & \text{for } \sigma_p^{}\neq \sigma_p^{'}
\end{cases}~,
\label{local commutation relation}
}
where $w[M_p^{}]$ is in general a positive functional defined on connected $p$-dimensional (open) surfaces $M_{p}^{}$.  
We assume that it is independent of the center-of-mass position of $M_p^{}$ and  satisfies the fusion property 
\aln{w[M_p^{}]w[M_p^{}]=w[M_p^{}+M_p^{'}]~,
\label{omega fusion}
}
where $M_p^{}+M_p^{'}$ denotes the fusion of surfaces as illustrated in Fig.~\ref{fig:fusion}.  
As for the normalization, we fix it by\footnote{
This condition can be equivalently written as  
\aln{
\sum_{\substack{C_{p}^{}~{\rm s.t.}\\ C_p^{}\pm \partial \sigma_{p+1}^{}\in \Gamma_p^{}}}w[C_p^{}]=1~,
}
by choosing a specific $\sigma_{p+1}^{}$ in $\Sigma_{D-1}^{}$. 
} 
\aln{
\sum_{C_p^{}\in \Gamma_p^{}}
w[C_p^{}]=\sum_{x\in \Lambda_{D-1}^{}}1~,
\label{concrete w condition}
}
where $x=(x^1,\cdots,x^{D-1})$ denotes a site on $\Lambda_{D-1}^{}$, and $\Gamma_p^{}$ is the set of all embeddings of $S^p$ in $\Lambda_{D-1}^{}$. 
A simplest example is 
\aln{w[C_p^{}]=\frac{1}{\#\{\text{embeddings with $x=0$}\}}~,
\label{simplest weight}
}
where $\#\{\text{embeddings with $x=0$}\}$ denotes the total number of embedded surfaces $C_p^{}$ with a fixed center-of-mass position, $x=0$. 
In the continuum limit, Eq.~(\ref{concrete w condition}) becomes  
\aln{
\int {\cal D}Xw[C_p^{}]=\int_{\Sigma_{D-1}^{}} \star 1~. \nonumber 
\label{normalization of w}
} 
We will see that this normalization condition is required in order to obtain meaningful thermodynamic limits.  

Now, a surface operator can be naturally defined by 
\aln{
\hat{\phi}[C_p^{}]\coloneq \prod_{\sigma_p^{}\in C_p^{}}\hat{\phi}(\sigma_p^{})~,
}
which obeys the commutation relation
\aln{
[\hat{\phi}[C_p^{}],\hat{\phi}^\dagger[C_p^{'}]]=w[C_p^{}]^{-1}\delta_{C_p^{},C_p^{'}}^{}=\begin{cases} 1 & \text{for } C_p^{}=C_p^{'}
\\
0 & \text{for } C_p^{}\neq C_p^{'}
\end{cases}
\label{brane field commutation relation}
}
due to Eq.~(\ref{local commutation relation}) and (\ref{omega fusion}).   
Here, $w[C_p^{}]^{-1}$ can be interpreted as a functional counterpart of the metric determinant in Eq.~(\ref{bosonic commutation}). 
Using this commutation relation, one can also check Eq.~(\ref{p-form symmetry}), where the $p$-form conserved charge is explicitly given by
\aln{
\hat{Q}_p^{}[\Sigma_{D-p-1}^{}]=\sum_{C_p^{}\in \Gamma_p^{}}
w[C_p^{}]\mathrm{I}[C_p^{},\Sigma_{D-p-1}^{}]\hat{\phi}^\dagger [C_p^{}]\hat{\phi}[C_p^{}]~,
} 
As a consistency check, one can see that this reproduces the $0$-form charge~(\ref{0-form charge}) by putting $p=0$ and $\omega[C_p^{}]=\sqrt{g(x)}$ since $\mathrm{I}[x,\Sigma_{D-1}^{}]=1$. 

\begin{figure}
    \centering
    \includegraphics[scale=0.7]{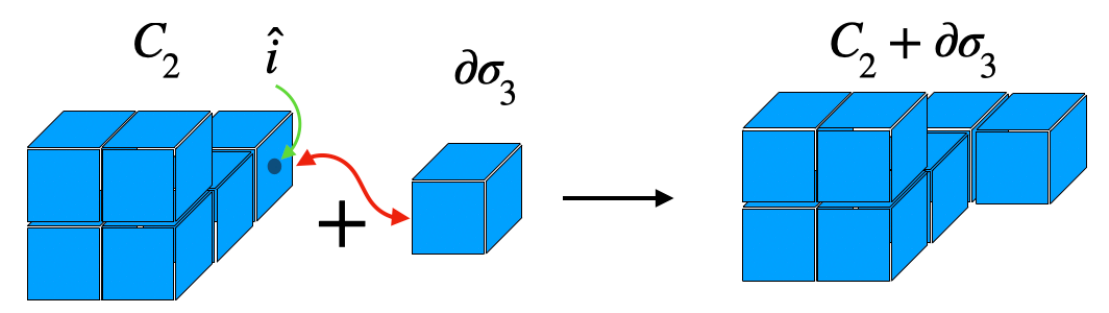}
    \caption{An infinitesimal deformation of a surface $C_2^{}$ in a hypercubic lattice.  
    }
    \label{fig:deformation1}
\end{figure}
In this setup, an infinitesimal variation of a general functional $\Psi[C_p^{}]$ is obtained by gluing a minimum $p$-dimensional closed surface $\partial \sigma_{p+1}^{}$ as 
\aln{
\label{functional variation}
\delta_{\hat{i}} \Psi[C_p^{}]\coloneq \Psi[C_p^{}+\partial \sigma_{p+1}^{}]-\Psi[C_p^{}]~,
}
where $\hat{i}$ denotes the center-of-mass position of $\sigma_p^{}\in C_p^{}$, at which $\partial \sigma_{p+1}^{}$ is glued.  
See Fig.~\ref{fig:deformation1} for an illustration.   
Using this, we can define an Hamiltonian similar to Eq.~(\ref{particle Hamiltonian}) as 
\aln{
\hat{H}=\sum_{C_p^{}\in \Gamma_p^{}}
w[C_p^{}]\left\{\frac{1}{g^{}\mathrm{Vol}[C_p^{}]}\sum_{\hat{i}\in C_p^{}}\delta_{\hat{i}}^{}\hat{\phi}^\dagger[C_p^{}]\delta_{\hat{i}}^{} \hat{\phi}[C_p^{}]
+V[C_p^{}]\hat{\phi}^\dagger[C_p^{}]\hat{\phi}[C_p^{}]
+U(\hat{\phi}^\dagger \hat{\phi})
\right\}~,
\label{surface hamiltonian}
}
where $g$ is a (dimensionful) coupling, and $V[C_p^{}]$ is an external potential.\footnote{
For simplicity, we assume that $U(\hat{\phi}^\dagger\hat{\phi})$ does not depend on  $C_p^{}$ explicitly.  
}
In addition to the $p$-form global symmetry, this Hamiltonian (accidentally) possesses an $\mathrm{U}(1)$ $0$-form global symmetry, whose conserved charge is 
\aln{
\hat{Q}^{}=\sum_{C_p^{}\in \Gamma_p^{}}w[C_p^{}]\hat{\phi}^\dagger [C_p^{}]\hat{\phi}[C_p^{}]~,
}
which corresponds to the total number of surfaces embedded in $\Sigma_{D-1}^{}=\Lambda_{D-1}^{}$.  

In this model, the variational state is simply given by 
\aln{|\psi_G^{}\rangle=\sqrt{{\cal N}}\exp\left(\sum_{C_p^{}\in \Gamma_p^{}}w[C_p^{}]\psi_G^{}[C_p^{}]\hat{\phi}^\dagger [C_p^{}]
\right)|0\rangle~
\label{discrete variational state}
}
without the prescription.  
One can check the coherent property 
\aln{
\hat{\phi}[C_p^{}]|\psi_G^{}\rangle=\psi_G^{}[C_p^{}]|\psi_G^{}\rangle~
}
by using the commutation relation~(\ref{brane field commutation relation}). 
Taking the expectation value of the Hamiltonian~(\ref{surface hamiltonian}) by the variational state~(\ref{discrete variational state}), we obtain the free-energy functional 
\aln{
F[\psi_G^{};\Sigma_{D-p-1}^{}]=\sum_{C_p^{}\in \Gamma_p^{}}w [C_p^{}]&\left\{\frac{1}{g\mathrm{Vol}[C_p^{}]}\sum_{\hat{i}\in C_p^{}}\delta_{\hat{i}}^{}\psi_G^{}[C_p^{}]^*\delta_{\hat{i}}^{} \psi_G^{}[C_p^{}]
+V[C_p^{}]\psi_G^{}[C_p^{}]^*\psi_G^{}[C_p^{}]
+U(\psi_G^*\psi_G^{})
\right\}
\nn
&-\mu_p^{}\left(\langle \psi_G^{}|\hat{Q}_p^{}[\Sigma_{D-p-1}^{}]|\psi_G^{}\rangle-Q_p^{}\right)-\mu^{}\left(\langle \psi_G^{}|\hat{Q}|\psi_G^{}\rangle-Q^{}\right)~.
\label{surface energy functional}
}
Finally, by taking the variations with respect to $\psi_G^{}[C_p^{}]^*$ and chemical potentials, we obtain  
\aln{
&
0=\left(-\frac{1}{gw[C_p^{}]}\sum_{\hat{i}\in C_p^{}}\delta_{\hat{i}}\frac{w[C_p^{}]}{\mathrm{Vol}[C_p^{}]}\delta_{\hat{i}}+V[C_p^{}]-\mu_p^{}\mathrm{I}[C_p^{},\Sigma_{D-p-1}^{}]-\mu
\right)\psi_G^{}[C_p^{}]+\frac{\delta U(\psi^{*}_G\psi_G^{})}{\delta \psi_G^{}[C_p^{}]^*}~,
\label{surface variational equation}
\\
&\sum_{C_p^{}\in \Gamma_p^{}}w[C_p^{}]\mathrm{I}[C_p^{},\Sigma_{D-p-1}^{}]{\psi}_G^{} [C_p^{}]^*\psi_G^{}[C_p^{}]=Q_p^{}~,\quad \sum_{C_p^{}\in \Gamma_p^{}}w[C_p^{}]{\psi}_G^{} [C_p^{}]^*\psi_G^{}[C_p^{}]=Q~. 
} 
Equation~(\ref{surface variational equation}) can be regarded as a generalization of the conventional Gross-Pitaevskii equation~(\ref{GP equation}), and we refer to it as the generalized Gross-Pitaevskii equation throughout this paper. 

\

\noindent {\bf Continuum limit}\\
We can also study the (classical) continuum limit in this theory. 
To this end, we need to clarify the precise meaning of the infinitesimal variation~(\ref{functional variation}). 
In general, let us consider a functional field $\Psi[C_p^{}]$ such that $C_p^{}$ can be a connected {\it intersecting} surface.      
In this case, a minimum closed surface $\partial\sigma_{p+1}^{}$ can be glued to the original surface $C_p^{}$ in any $(p+1)$-dimensional subspaces of $\Lambda_{D-1}^{}$, provided that $\partial\sigma_{p+1}^{}\cap C_p^{}\neq \emptyset$.  
Namely, we allow that $C_p^{}+\partial \sigma_{p+1}^{}$ can be a connected intersecting surface even when $C_p^{}$ is an embedded surface of $S^p$.   
This deformation then leads to the functional derivative known as the {\it area derivative}~\cite{Iqbal:2021rkn,Hidaka:2023gwh,Kawana:2024fsn,Kawana:2024qmz,Kawana:2025vvf,Kawana:2025vbi}
\aln{
\label{area derivative}
\frac{\delta \Psi[C_p^{}]}{\delta \Sigma^{k_1^{}\dots k_{p+1}^{}}(\xi)}=\lim_{a\rightarrow \infty}\frac{\Psi[C_p^{}+\partial \sigma_{p+1}^{}]-\Psi[C_p^{}]}{a^{p+1}}~,
}
in the continuum limit, where $\xi=(\xi^1,\cdots,\xi^p)$ is the intrinsic coordinate on $C_p^{}$, representing the location at  which $\partial \sigma_{p+1}^{}$ is glued, $a^{p+1}$ is the volume of $\sigma_{p+1}^{}$, and the indices $(k_{1}^{},k_2^{},\dots,k_{p+1}^{})$ mean that $\partial \sigma_{p+1}^{}$ lies within the $(p+1)$-dimensional subspace spanned by the direction~$(k_{1}^{},k_2^{},\dots,k_{p+1}^{})$ in $\Lambda_{D-1}^{}$. 
Since a permutation of the target-space coordinates corresponds to a change of the orientation of $C_p^{}$, these area derivatives satisfy the anti-symmetric property
\aln{
\frac{\delta}{\delta \Sigma^{\sigma(k_1^{})\cdots \sigma(k_{p+1}^{})}(\xi)}=\text{sgn}(\sigma)\frac{\delta}{\delta \Sigma^{k_1^{}\cdots k_{p+1}^{}}(\xi)}
}
for $^\forall \sigma\in S_{p+1}^{}$.  

We now return to Eq.~(\ref{functional variation}).    
Since we assume that $C_p^{}$ is an embedding of $S^p$ (i.e. self-avoiding surface), $\partial \sigma_{p+1}^{}$ cannot be attached to $C_p^{}$ in an arbitrary manner, but has to share at least one $p$-dimensional hypercube $\sigma_p^{}$ on $C_p^{}$, as illustrated in Fig.~\ref{fig:deformation1} for $p=2$. 
%
%
This implies that Eq.~(\ref{functional variation}) corresponds to a     projected area derivative as~\cite{Kawana:2025vbi}
\aln{
\frac{\delta}{\delta \Sigma^{k}(\xi)}&\coloneq \lim_{a\rightarrow \infty}\frac{\delta_{\hat{i}}\Psi[C_p^{}]}{a^{p+1}}=\frac{1}{p!}\sum_{l_1^{},\cdots,l_p^{}}(E_p^{}(\xi))^{l_1^{}\cdots l_{p}^{}}\frac{\delta}{\delta \Sigma^{k l_1^{}\cdots l_p^{}}(\xi)}~,
\label{projected area derivative}
}
where 
\aln{
(E_p^{}(\xi))^{l_1^{}\cdots l_{p}^{}}=\frac{1}{\sqrt{h(\xi)}}\{X^{l_1^{}},\cdots,X^{l_p^{}}\}~
}
%
%
is the normalized volume element of $\sigma_p^{}\in C_p^{}$. 
Here,
\aln{
\{X^{l_1^{}},\cdots,X^{l_p^{}}\}=\epsilon^{k_1^{}\dots k_p^{}}\frac{\partial X^{l_1^{}}(\xi)}{\partial \xi^{k_1^{}}}\cdots \frac{\partial X^{l_p^{}}(\xi)}{\partial \xi^{k_p^{}}}
}
is the generalized Nambu-bracket~\cite{Nambu:1973qe} and 
\aln{
h(\xi)=\frac{1}{p!}\{X^{l_1^{}},\cdots,X^{l_p^{}}\}\{X_{l_1^{}},\cdots,X_{l_p^{}}\}
}
is the determinant of the induced metric on $C_p^{}$.

Now, using the projected area derivatives, the continuum limit of Eq.~(\ref{surface energy functional}) is given by  
\aln{
\label{continuum surface energy}
E[\psi_G^{}]=\int{\cal D}Xw[C_p^{}]\bigg\{
\int_{S^p}d^p\xi &\frac{\sqrt{h(\xi)}}{\mathrm{Vol}[C_p^{}]}\sum_{k=1}^{D-1}\frac{\delta \psi_G^{}[C_p^{}]^*}{\delta \Sigma^{k}(\xi)} \frac{\delta \psi_G^{}[C_p^{}]}{\delta \Sigma_{k}(\xi)}
+V[C_p^{}]\psi_G^{}[C_p^{}]^*\psi_G^{}[C_p^{}]+U(\psi_G^*\psi_G) 
\bigg\}~,
}
where the kinetic term is canonically normalized by the field redefinition. 
%
Correspondingly, the generalized Gross-Pitawvskii equation~(\ref{surface variational equation}) becomes
\aln{
0=\left(-\Box_p^{}+V[C_p^{}]-\mu_p^{}\mathrm{I}[C_p^{},\Sigma_{D-p-1}^{}]-\mu\right)\psi_G^{}[C_p^{}]+\frac{\delta U(\psi_G^{*}\psi_G^{})}{\delta \psi_G^{}[C_p^{}]^*}~,
\label{continuum surface variational equation}
} 
where 
\aln{
\Box_p^{}=\frac{1}{w[C_p^{}]}\int_{S^p} d^p\xi \sum_{k=1}^{D-1}\frac{\delta}{\delta \Sigma^{k}(\xi)}\frac{w[C_p^{}]\sqrt{h(\xi)}}{\mathrm{Vol}[C_p^{}]}\frac{\delta}{\delta \Sigma_{k}(\xi)}
\label{d'Alembert operator}
}
is a generalized Laplace operator. 
%

Lastly, we can also obtain the action leading to the Hamiltonian~(\ref{surface hamiltonian}) and the commutation relation~(\ref{brane field commutation relation}) as 
\aln{
S[\phi]=&\int_{t_0^{}}^{t_1^{}}dt\int {\cal D}X w[C_p^{}]\bigg\{i\left(\phi^*\dot{\phi}-\phi \dot{\phi}^*\right)-\int_{S^p}d^p\xi
\frac{\sqrt{h(\xi)}}{\mathrm{Vol}[C_p^{}]}\sum_{k=1}^{D-1}\frac{\delta \phi^*}{\delta \Sigma^{k}(\xi)} \frac{\delta \phi}{\delta \Sigma_{k}(\xi)}-V[C_p^{}]\phi^*\phi-U(\phi^*\phi)
\bigg\}~,
\label{non-relativistic brane theory}
}
where $\phi=\phi[t,C_p^{}]$.  
%
In the following discussion, we adopt the convention  
\aln{\int [dC_p^{}]\coloneq \int {\cal D}Xw[C_p^{}]^{}
} 
in order to simplify various path-integral expressions.

\subsection{More on surface interactions}

Since the content of this subsection is not essential for the following discussion, readers may skip this part on a first reading.  

Although we have formulated the variational method in a similar manner to conventional bosonic systems, several important issues   have been intentionally omitted in order to keep our formulation as simple as possible.    
%
First, surfaces can split and merge without violating the $\mathrm{U}(1)$ higher-form global symmetry. 
For example, we can consider the following nonlocal interaction
\aln{\label{topological interaction}
\int [dC_p^1]\int [dC_p^2]\int [dC_p^3]\delta(C_p^{1}-C_p^2-C_p^3)\hat{\phi}^\dagger[C_p^1]\hat{\phi}[C_p^2]\hat{\phi}[C_p^3]+{\rm h.c.}~,
}
where $\delta(C_p^{}-C_p^{'})$ is the continuum version of the delta function~(\ref{brane field commutation relation}). 
%
%
This interaction represents a merging of two surfaces $C_p^2,C_p^3$ into another surface $C_p^1$ and is invariant under the $p$-form global transformation~(\ref{p-form symmetry}) as 
\aln{&\delta(C_p^{1}-C_p^2-C_p^3)\hat{\phi}^\dagger[C_p^1]\hat{\phi}[C_p^2]\hat{\phi}[C_p^3]
\nn
\rightarrow \quad &\delta(C_p^{1}-C_p^2-C_p^3)e^{i\theta\int_{C_p^{2}+C_p^3-C_p^1}\Lambda_p^{}}\hat{\phi}^\dagger[C_p^1]\hat{\phi}[C_p^2]\hat{\phi}[C_p^3]=\delta(C_p^{1}-C_p^2-C_p^3)\hat{\phi}^\dagger[C_p^1]\hat{\phi}[C_p^2]\hat{\phi}[C_p^3]~.
}
On the other hand, this interaction explicitly breaks $\mathrm{U}(1)^{[0]}$, namely, the total number of surfaces is not conserved when this interaction exists.    
%
 
%
%
Second, it is possible to promote $\mathrm{U}(1)^{[0]}$ to a general (non-abelian) group $G^{[0]}$ by assigning a nontrivial representation $R$ of $G^{[0]}$ to the surface operator, $\hat{\phi}[C_p^{}]\rightarrow \hat{\phi}_a^{}[C_p^{}]$, where $a=1,2,\cdots,\mathrm{dim}R$ denotes the component of the representation $R$. 
However, invariance under such a nontrivial group $G^{[0]}$ is typically incompatible with the interactions such as Eq.~(\ref{topological interaction}) because they cannot be expressed in a $G^{[0]}$ invariant manner.    
%
%
%
In other words, one typically needs to break $G^{[0]}$ in order to introduce interactions describing splitting and merging of surfaces.

\begin{figure}
    \centering
    \includegraphics[scale=0.8]{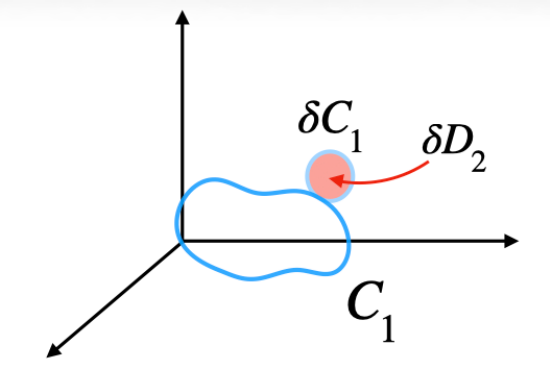}
    \caption{A surface deformation for $p=1$. 
    The red region is the interior region of $\delta C_1^{}$. 
    }
    \label{fig:deformation}
\end{figure}
Third, a microscopic Hamiltonian such as Eq.~(\ref{surface hamiltonian}) does not capture the dynamics of the center-of-mass motion of $p$-dimensional surface $C_p^{}$. 
%
%
%
This can be explicitly seen by noting that an infinitesimal volume element $\delta \Sigma^{k_1^{}\cdots k_{p+1}^{}}$ can be expressed as~\cite{Hidaka:2023gwh,Kawana:2024fsn,Kawana:2024qmz,Kawana:2025vvf,Kawana:2025vbi}
\aln{
\delta \Sigma^{k_1^{}\cdots k_{p+1}^{}}=\int_{\delta D_{p+1}^{}}dX^{k_1^{}}\wedge \cdots \wedge dX^{k_{p+1}^{}}~,
} 
where $\delta D_{p+1}^{}$ is the interior region of an infinitesimal  deformation $\delta C_p^{}$ as depicted in Fig.~\ref{fig:deformation}.  
By introducing the center-of-mass coordinates and the relative coordinates as 
\aln{\label{COM}
x^k &\coloneq \frac{1}{\mathrm{Vol}[C_p^{}]}\int_{S^p}d^p\xi \sqrt{h(\xi)}X^k(\xi)~,\quad  Y^k(\xi)\coloneq X^k(\xi)-x^k~,
}
one finds 
\aln{
\delta \Sigma^{k_1^{}\cdots k_{p+1}^{}}=\int_{\delta D_{p+1}^{}}dY^{k_1^{}}\wedge \cdots \wedge dY^{k_{p+1}^{}}~,
}
which shows that the area-volume elements are independent of the center-of-mass position for $p\geq 1$.  
In other words, the kinetic term in Eq.~(\ref{surface hamiltonian}) encodes   the functional variations associated only with the relative motion and does not involve the dynamics of the center-of-mass motion.  

Considering this fact, if one wishes to incorporate the dynamics of the center-of-mass motion as well, the Hamiltonian~(\ref{surface hamiltonian}) should be generalized as
\aln{
\hat{H}=\int [dC_p^{}]\bigg\{\frac{1}{m}\partial_k^{}\hat{\phi}_a^\dagger[C_p^{}]\partial_k^{}\hat{\phi}_a^{}[C_p^{}]&+\int_{S^p}d^p\xi\frac{\sqrt{h(\xi)}}{\mathrm{Vol}[C_p^{}]}\frac{\delta \hat{\phi}_a^\dagger [C_p^{}]}{\delta \Sigma_{k}(\xi)} \frac{\delta \hat{\phi}_a^{}[C_p^{}]}{\delta \Sigma^{k}(\xi)}
+V[C_p^{}]\hat{\phi}^\dagger_a[C_p^{}]\hat{\phi}_a^{}[C_p^{}]+U(\hat{\phi}^\dagger_a \hat{\phi}_a^{})
\bigg\}~,
\label{general surface Hamiltonian}
}
where $m^{}$ is a dimensioful constant, $\partial_k^{}=\partial/\partial x^k$, and $k$ runs over the spatial directions.\footnote{
For $p=0$, two kinetic terms coincide with each other since the area derivative reduces to he ordinary derivative, $\frac{\delta}{\delta \sigma^k}=\frac{\partial}{\partial x^k}$.  
}
This is a generic Hamiltonian of closed surface that possesses a global symmetry $G^{[0]}\times \mathrm{U}(1)^{[p]}$ without nonlocal interactions such as Eq.~(\ref{topological interaction}). 
In the vanishing volume limit $\mathrm{Vol}[C_p^{}]\rightarrow 0$, Eq.~(\ref{general surface Hamiltonian}) reduces to the particle Hamiltonian~(\ref{particle Hamiltonian}) since the second kinetic term  vanishes.     
%

\section{Mean field analysis}\label{mean field analysis}

In this section, we perform a mean-field analysis based on the generalized Gross-Pitaevskii equation~(\ref{continuum surface variational equation}). 
See also Refs.~\cite{Hidaka:2023gwh,Kawana:2024fsn,Kawana:2024qmz,Kawana:2025vvf,Kawana:2025vbi} for similar studies.      
In this section, we consider $\Sigma_{D-1}^{}=\mathbb{R}^{D-1}$ except in Section~\ref{discrete symmetry}. 

%

\subsection{Uniform bosonic surfaces}
%
When the external potential $V[C_p^{}]$ is absent, the variational equation~(\ref{continuum surface variational equation}) becomes  
\aln{
0=(-\Box_p^{}-\mu)\psi_G^{}[C_p^{}]+\frac{\delta U(\psi_G^*\psi_G^{})}{\delta \psi_G^{}[C_p^{}]^*}~,
}
which has the uniform solution  
\aln{
\mu=\frac{1}{\psi_G^{}[C_p^{}]}\frac{\delta U(\psi_G^*\psi_G^{})}{\delta \psi_G^{}[C_p^{}]^*}~,
}
when the potential $U(\psi_G^*\psi_G)$ is bounded below. 
For example, the standard quartic interaction $U(\psi_G^*\psi_G)=\lambda (\psi_G^*\psi)^2/2~,~\lambda>0$ leads to 
\aln{n\coloneq |\psi_G^{}[C_p^{}]|^2=\frac{\mu}{\lambda}~,
\label{particle density}
}
which is related to the total number of surfaces as 
\aln{Q=\int [dC_p^{}]|\psi_G^{}[C_p^{}]|^2=V_{D-1}^{}n~,
}
where $V_{D-1}^{}$ is a regularized volume of $\mathbb{R}^{D-1}$ and we have used the normalization condition~(\ref{normalization of w}).  
%
The corresponding ground state is written as  
\aln{
|\psi_G^{}\rangle=\sqrt{{\cal N}}\exp\left(\sqrt{n}\int[dC_p^{}]\hat{\phi}[C_p^{}]\right)|0\rangle~,
}
which is a superposition of all embedded surfaces and can be interpreted as a surface generalization of the string condensation in topologically ordered phases in $(1+2)$ dimension~\cite{Levin:2004mi}. 
By construction, we have
\aln{\langle \psi_G^{}|\hat{\phi}[C_p^{}]|\psi_G^{}\rangle=\sqrt{n}\neq 0\quad \text{for all $C_p^{}\in \Gamma_p^{}$}~,
}  
which implies that the $\mathrm{U}(1)$ $p$-form global symmetry is spontaneously broken. 

\

\noindent {\bf NON-CONDENSED PHASE}\\
It is also instructive to consider $\mu\leq 0$. 
As long as the interaction potential $U(X)$ is an monotonically increasing function of $X$, the uniform solution is $\psi_G^{}[C_p^{}]=0$, and the corresponding ground state is trivial, $|\psi_G^{}\rangle=|0\rangle$. 
In this case, we have  
\aln{\langle \psi_G^{}|\hat{\phi}[C_p^{}]|\psi_G^{}\rangle=0\quad \text{for all $C_p^{}\in \Gamma_p^{}$}~,
\label{zero VEV}
}
i.e., the $\mathrm{U}(1)$ $p$-form global symmetry is unbroken.    

At this point, recall that the presence or absence of condensation is characterized by the long-distance behavior of the two-point correlation function 
\aln{\label{two point function}
G^{}(x)\coloneq \mathrm{Tr}(\hat{\rho}\hat{\phi}^\dagger(x)\hat{\phi}(0))~
} 
in conventional bosonic systems, where $\hat{\rho}$ denotes a general density matrix. 
By substituting the mode expansion~(\ref{function expansion}), this correlation function is written as 
\aln{
=\sum_{i,j}\psi_i^{*}(x)\psi_j^{}(0){\rm Tr}(\hat{\rho}\hat{a}_i^\dagger \hat{a}_j^{})~.
} 
When the density matrix $\hat{\rho}$ represents a uniform equilibrium ensemble such as the grand-canonical ensemble, we have
\aln{{\rm Tr}(\hat{\rho}\hat{a}_i^\dagger \hat{a}_j^{})=\delta_{ij}^{}{\rm Tr}(\hat{\rho}\hat{a}_i^\dagger \hat{a}_i^{})\equiv \delta_{ij}^{}\langle \hat{N}_i^{}\rangle~,
}
and Eq.~(\ref{two point function}) becomes
\aln{&=\sum_{i}\langle \hat{N}_i^{}\rangle\psi_i^{*}(x)\psi_i^{}(0)
=\frac{\langle \hat{N}_0^{}\rangle}{Q}n+\sum_{k\neq 0}\frac{\langle \hat{N}_i^{}\rangle}{V_{D-1}^{}}e^{-ik\cdot x}~,
}
where we have assumed a uniform bose gas, and $n$ is the particle number density at zero temperature such as Eq.~(\ref{particle density}). 
Bose-Einstein condensation is realized when $\langle\hat{N}_0^{}\rangle\propto Q$, which then implies 
\aln{
\lim_{|x|\rightarrow \infty}G(x)=\frac{\langle \hat{N}_0^{}\rangle}{Q}n<\infty~
}
by the Riemann-Lebesgue theorem. 
This is the off-diagonal longe-range order and is widely accepted as the criterion for Bose-Einstein condensation~\cite{RevModPhys.71.463,RevModPhys.73.307,Pethick_Smith_2008,Griffin_Nikuni_Zaremba_2009,Yukalov:2011qj,Proukakis}.  
When the system is not in the condensed phase, the long-distance behavior of the correlation function is typically
\aln{G(x)\approx  \int\frac{d^{D-1}k}{(2\pi)^{D-1}}\langle N_k^{}\rangle e^{-ik\cdot x}\propto e^{-f(|x|/\xi)}~,
\label{damping behavior}
} 
where $f(y)$ is a smooth monotonically increasing function and $\xi$ is a  correlation length. 

Now let us view  this criterion from the perspective of higher-form global  symmetries. 
A key observation is that the bilocal operator $\hat{\phi}^\dagger(x)\hat{\phi}(y)$ can be regarded as a closed surface operator by regarding the two end points, $x$ and $y$, as boundaries of a straight line $L_1^{}$, i.e.,
\aln{
\partial L_1^{}=\{x,y\}~.
\label{line boundary}
} 
In this sense, the two-point correlation function~(\ref{two point function}) is an expectation value of the closed surface operator $\hat{\phi}^\dagger(x)\hat{\phi}(y)$, which exhibits an exponential behavior~(\ref{damping behavior}) in the non-condensed phase. 
In the case of $p$-form symmetry, the counterpart of Eq.~(\ref{line boundary}) is 
\aln{\partial L_{p+1}^{}=C_p^{}~,
}
where $L_{p+1}^{}$ is the $(p+1)$-dimensional minimal surface bounded by $C_p^{}$.  
This highlights a crucial difference between $0$-form and higher-form cases. 
In the higher-form case, the boundary consists of a single closed surface $C_p^{}$, which implies that the counterpart of $\hat{\phi}^\dagger(x)\hat{\phi}(y)$ is the surface operator $\hat{\phi}[C_p^{}]$ itself.     
%
Accordingly, condensation of surfaces is characterized by the asymptotic behavior of the expectation value 
\aln{
\langle \hat{\phi}^{}[C_p^{}]\rangle\coloneq \mathrm{Tr}(\hat{\rho}\hat{\phi}[C_p^{}])~,
\label{general tad pole}
}
for $\mathrm{Vol}[L_{p+1}^{}]\rightarrow \infty$.  
As long as we focus on the ground state, $\hat{\rho}=|\psi_G^{}\rangle\langle \psi_G^{}|$, this expectation value trivially vanishes as in Eq.~(\ref{zero VEV}).       
At a finite temperature, however, there can exist small corrections, analogous to Eq.~(\ref{damping behavior}), for $\mathrm{Vol}[L_{p+1}^{}]\rightarrow \infty$. 
To illustrate this, let us consider the grand-canonical ensemble: 
\aln{\hat{\rho}_{\rm eq}^{}=\frac{e^{-\beta \hat{H}+\mu \hat{Q}}}{Z}~
}   
as a concrete example. 
By following the standard path-integral technique, the expectation value  can be expressed in a path-integral form as
\aln{
\langle \hat{\phi}^{}[C_p^{}]\rangle_\beta^{}=\frac{1}{Z}\int {\cal D}\phi^*{\cal D}\phi~\phi[0,C_p^{}]e^{-S_E^{}[\phi;\beta]}~,
\label{tad pole in finite}
} 
where 
\aln{
S_E^{}[\phi;\beta]=&\int_0^{\beta}d\tau \int [dC_p^{}]\bigg\{\phi^* \partial_\tau^{}\phi+\int_{S^p}d^p\xi
\frac{\sqrt{h(\xi)}}{\mathrm{Vol}[C_p^{}]}\sum_{k=1}^{D-1}\frac{\delta \phi^*}{\delta \Sigma^{k}(\xi)} \frac{\delta \phi}{\delta \Sigma_{k}(\xi)}-\mu\phi^*\phi +U(\phi^*\phi)
\bigg\}~
}
is the Euclidean action, and we have set $V[C_p^{}]=0$ for simplicity.   
In general, the path-integral is dominated by saddle-point configurations determined by
\aln{
0=\partial_\tau^{}\phi-(\Box_p^{}+\mu)\phi+\frac{\delta U(\phi^*\phi)}{\delta \phi}~.
\label{saddle point equation}
}
In the following, we focus on static solutions, $\phi=\phi[C_p^{}]$ for simplicity.  
The trivial solution is, of course, the ground state $\phi=0$, which contributes to the Euclidean action as 
\aln{
S_E^{}[0,\tau]=\int_0^{\beta}d\tau \int_{\Sigma_{D-1}^{}}{\cal D}X w[C_p^{}]U(0)=\beta V_{D-1}^{}U(0)~.
}  
This is nothing but the vacuum-energy contribution. 
Note that we have used the normalization condition~(\ref{normalization of w}) again.    
Can we find other nontrivial saddles ?       
The finiteness of action density requires the boundary condition 
\aln{
\phi\rightarrow 0\quad \text{for $\mathrm{Vol}[L_{p+1}^{}]\rightarrow \infty$}
} 
which in turn leads to the approximate saddle-point equation
\aln{
0\approx (\Box_p^{}+\mu)\phi[C_p^{}]
\label{generalized wave equation}
}
for $\mathrm{Vol}[L_{p+1}^{}]\rightarrow \infty$. 
This is a generalized wave equation, where the Laplace operator $\Box_p^{}$ is defined in terms of the functional derivatives, as given in Eq.~(\ref{d'Alembert operator}). 
Even in this simplified form, solving this equation in full generality  is a highly nontrivial problem.      
Below, we present one concrete solution that corresponds to the exponential behavior $G(x)\propto e^{-|x|/\xi}$ in the particle case.  

As an ansatz, let us consider 
\aln{\phi[C_p^{}]=f(\mathrm{Vol}[L_{p+1}^{}])~,\quad f(z)\in \mathbb{R}~, 
}
where $f(x)$ is supposed to satisfy the boundary condition $f(\infty)=0$.  
Substituting this ansatz into Eq.~(\ref{generalized wave equation}), we obtain 
\aln{p(z)f''(z)-q(z)f'(z)-|\mu|f(z)\approx 0~,\quad z=\mathrm{Vol}[L_{p+1}^{}]~,
\label{simplified eom}
}
where 
\aln{p(z)&=\int_{S^p}d^p\xi\frac{\sqrt{h(\xi)}}{\mathrm{Vol}[C_p^{}]}\left[\frac{1}{p!}(E_{p+1}^{}(\xi))^{k k_1^{}\cdots k_{p}^{}}(E_{p}^{}(\xi))_{k_1^{}\cdots k_{p}^{}}\right]^2~,
\\
q(z)&=\frac{1}{w[C_p^{}]}\int_{S^p}d^p\xi\frac{\sqrt{h(\xi)}}{\mathrm{Vol}[C_p^{}]} \frac{1}{p!}\frac{\delta [w[C_p^{}](E_{p+1}^{}(\xi))^{k k_1^{}\cdots k_{p}^{}}(E_{p}^{}(\xi))_{k_1^{}\cdots k_{p}^{}}]}{\delta \Sigma^{\mu}(\xi)}~.
}
Here, $(E_{p+1}^{}(\xi))_{k_1^{}\cdots k_{p+1}^{}}$ is the normalized volume element of the minimal surface $L_{p+1}^{}$, and we have used~\cite{Hidaka:2023gwh,Kawana:2024fsn,Kawana:2024qmz,Kawana:2025vvf}
\aln{\frac{\delta \mathrm{Vol}[L_{p+1}^{}]}{\delta \Sigma^{k_1^{}\cdots k_{p+1}^{}}(\xi)}=(E_{p+1}^{}(\xi))_{k_1^{}\cdots k_{p+1}^{}}~.
}   
Since we are interested in topological aspects of the solution, it is sufficient to consider a spherically symmetric surface $C_p^{}\simeq S^p$ characterized by a radius $r$, and embedded in the $(p+1)$-dimensional subspace with the polar coordinates $(r,\xi_1^{},\cdots,\xi_p^{})$.     
In this case, the weight functional reduces to an ordinary function,  $w[C_p^{}]=w(r)$, and the normalized volume elements are explicitly given by  
\aln{(E_p^{}(\xi))^{k_1^{}\cdots k_p^{}}&=\frac{\epsilon^{k_1^{}\cdots k_p^{}}}{\sqrt{h(\xi)}}~,\quad k_i^{}\in \{\xi_1^{},\cdots,\xi_p^{}\}~,
\\
(E_{p+1}^{}(\xi))^{k_1^{}\cdots k_{p+1}^{}}&=\frac{\epsilon^{k_1^{}\cdots k_{p+1}^{}}}{\sqrt{h(\xi)}}~,\quad k_i^{}\in \{r,\xi_1^{},\cdots,\xi_p^{}\}~,
}
by which we obtain 
\aln{
p(z)=1~,\quad q(z)=\frac{1}{r^p}\frac{d\log w(r)}{\partial r}~.
}
This implies that the equation of motion~(\ref{simplified eom}) is approximated by 
\aln{
f''(z)-|\mu|f(z)\approx 0~
\label{simplified EOM}
}
for $z\rightarrow \infty$, and we obtain the area-law behaviour  
\aln{
f(z)\propto e^{-|\mu|z}=e^{-|\mu|\mathrm{Vol}[L_{p+1}^{}]}~.  
\label{Area law}
}
In particular, one can see that this reproduces the large-distance behavior of the two-point function $G(x)\sim e^{-|x|/\xi}$ by formally putting $p=0$. 
Summarizing all, there are at least two saddle-point solutions that contribute to the expectation value~(\ref{tad pole in finite}) as 
\aln{
\langle \hat{\phi}[C_p^{}]\rangle_\beta^{}=0+c(\beta,\mu)e^{-|\mu|\mathrm{Vol}[C_p^{}]}+\cdots ~,
}
where $c(\beta,\mu)$ denotes an undetermined relative weight between the  two saddle-point solutions.     
Although we cannot exclude the existence of other saddle-point solutions at this stage, the criterion for condensation shall be now stated as follows: In the condensed phase, the expectation value of the surface operator $\hat{\phi}[C_p^{}]$ approaches a finite constant for $\mathrm{Vol}[L_{p+1}^{}]\rightarrow \infty$~(i.e., the perimeter law), whereas it obeys the area law in the non-condensed phase.   
See also Refs.~\cite{Lake:2018dqm,Iqbal:2021rkn,Hidaka:2023gwh} and references therein for the related studies.

\subsection{Low energy excitations}
In ordinary interacting bosonic systems, low-energy excitations can be analyzed by expanding a given Hamiltonian in terms of the fluctuation $\delta \hat{\phi}(x)=\hat{\phi}(x)-\psi_G^{}(x)$ and diagonalizing the second order terms.   
Then, we typically find gapless sound-wave modes whose dispersion relation is $\varepsilon_k^{}\simeq v|k|$, where $v\geq 0$ is the sound velocity.   

In the present theory of interacting surfaces, however, it is challenging to carry out a similar Bogoliubov analysis in a systematic manner due to the functional nature of the field $\hat{\phi}[C_p^{}]$.     
Nevertheless, the presence or absence of gapless modes can be still analyzed by focusing on specific functional degrees of freedom.  
For instance, a natural candidate for low-energy gapless excitation is given by the phase modulation\footnote{Here, we use $\psi[C_p^{}]$ instead of $\psi_G^{}[C_p^{}]$ to emphasize that it represents a general excitation rather than the ground state.  
}
\aln{
\psi^{}[C_p^{}]=\sqrt{n}\times \exp\left(i\int_{C_p^{}}A_p^{}\right)~,
\label{phase modulation}
}
where $A_p^{}$ is a $p$-form field. 
When $G^{[p]}=\mathrm{U}(1)^{[p]}$, only the kinetic term in Eq.~(\ref{continuum surface energy}) is relevant for the low-energy dynamics of $A_p^{}$.  
Substituting Eq.~(\ref{phase modulation}) into the kinetic term, we have 
\aln{
\frac{n}{2(p!)^2}\int [dC_p^{}]\int_{S^p}d^p\xi  \frac{\sqrt{h(\xi)}}{\mathrm{Vol}[C_p^{}]}\{F_{k\nu_1^{}\cdots \nu_p^{}}^{}(X(\xi))(E_p^{}(\xi))^{\nu_1^{}\cdots \nu_p^{}}\}\{{F^k}_{\rho_1^{}\cdots \rho_p^{}}^{}(X(\xi))(E_p^{}(\xi))^{\rho_1^{}\cdots \rho_p^{}}\}~,
\label{kinetic term of A}
} 
where $F_{p+1}^{}=dA_p^{}=\frac{1}{(p+1)!}F_{\mu_1^{}\cdots \mu_{p+1}}^{}dX^{\mu_1^{}}\wedge\cdots \wedge dX^{\mu_{p+1}^{}}$ is the (spatial) field strength. 
Using the center-of-mass and relative coordinates~(\ref{COM}), the integrand in Eq.~(\ref{kinetic term of A}) can be expressed as 
\aln{
F_{k\nu_1^{}\cdots \nu_p^{}}^{}(X(\xi)){F^k}_{\rho_1^{}\cdots \rho_p^{}}^{}(X(\xi))=\int \frac{d^{D-1}k}{(2\pi)^{D-1}}e^{ik\cdot (x+Y)}
{\tilde{F}}_{k\nu_1^{}\cdots \nu_p^{}}^{}(k){{\tilde{F}^k}}_{\rho_1^{}\cdots \rho_p^{}}(k)~,
}
where $\tilde{F}_{\mu_1^{}\cdots \mu_{p+1}^{}}^{}(k)$ is the Fourier mode of $F_{\mu_1^{}\cdots \mu_{p+1}^{}}^{}(X)$.  
By putting this into Eq.~(\ref{kinetic term of A}), we obtain
\aln{
\int d^{D-1}k\left(\int \frac{d^{D-1}x}{(2\pi)^{D-1}}~e^{ik\cdot x}\right)
&{\tilde{F}}_{k\nu_1^{}\cdots \nu_p^{}}^{}(k){{\tilde{F}^k}}_{\rho_1^{}\cdots \rho_p^{}}(k)
\nn
\times &\int {\cal D}Yw[C_p^{}]\int_{S^p}d^p\xi \frac{\sqrt{h(\xi)}}{\mathrm{Vol}[C_p^{}]}
e^{ik\cdot Y(\xi)}(E_p^{}(\xi))^{\nu_1^{}\cdots \nu_p^{}}(E_p^{}(\xi))^{\rho_1^{}\cdots \rho_p^{}}
\nn
={\tilde{F}}_{\mu\nu_1^{}\cdots \nu_p^{}}^{}(0){{\tilde{F}^\mu}}_{\rho_1^{}\cdots \rho_p^{}}(0)
&\int_{S^p}d^p\xi \frac{\sqrt{h(\xi)}}{\mathrm{Vol}[C_p^{}]}\langle (E_p^{}(\xi))^{\nu_1^{}\cdots \nu_p^{}}(E_p^{}(\xi))^{\rho_1^{}\cdots \rho_p^{}}\rangle~,
\label{kinetic term deformation}
}
where 
\aln{
\langle (E_p^{}(\xi))^{\nu_1^{}\cdots \nu_p^{}}(E_p^{}(\xi))^{\rho_1^{}\cdots \rho_p^{}}\rangle=\int {\cal D}Yw[C_p^{}](E_p^{}(\xi))^{\nu_1^{}\cdots \nu_p^{}}(Y(\xi))(E_p^{}(\xi))^{\rho_1^{}\cdots \rho_p^{}}~.
}
%
As long as space(time) symmetries are not broken, this expectation value is 
\aln{
\langle (E_p^{}(\xi))^{\nu_1^{}\cdots \nu_p^{}}(E_p^{}(\xi))^{\rho_1^{}\cdots \rho_p^{}}\rangle &=\frac{1}{(D-1)^p}\sum_{\sigma,\sigma'\in S_p^{}}\mathrm{sgn}(\sigma)\mathrm{sgn}(\sigma')\delta^{\sigma(\nu_1^{})\sigma'(\rho_1^{})}\cdots \delta^{\sigma(\nu_p^{})\sigma'(\rho_p^{})}~,
\label{k expansion}
} 
which satisfies the normalization $\langle (E_p^{}(\xi))^{\nu_1^{}\cdots \nu_p^{}}(E_p^{}(\xi))_{\nu_1^{}\cdots \nu_p^{}}\rangle=p!$.  
Substituting this into Eq.~(\ref{kinetic term deformation}), we obtain the effective energy functional of $A_p^{}$:
\aln{\label{Maxwell theory}
E[A_p^{}]=\frac{n}{(D-1)^p}\tilde{F}_{k_1^{}\cdots k_{p+1}^{}}(0)\tilde{F}^{k_1^{}\cdots k_{p+1}^{}}(0)
\propto n\int_{\Sigma_{D-1}^{}} F_{p+1}^{}\wedge \star F_{p+1}^{}~. 
} 
which describes a gapless $p$-form field.  
%
%

How about other fluctuations ? 
In general, the fluctuation $\psi[C_p^{}]-\psi_G^{}[C_p^{}]$ contains infinitely many local fields as in string field theory. 
For example, one can also consider an expansion of the phase with respect to  $\partial X^k(\xi)/\partial \xi^i$ as 
\aln{
\psi[C_p^{}]&=\sqrt{n}\times \exp\left(i\int_{C_p^{}}A_p^{}+i\int_{C_p^{}}d^p\xi\sqrt{h}\left\{S(X(\xi))+h^{ij}(\xi)\frac{\partial X^k}{\partial\xi^i}\frac{\partial X^l}{\partial\xi^j}H_{kl}^{}(X(\xi))+\cdots\right\}\right)~,
}
where $S(X)$ is a scalar field and $H_{\mu\nu}^{}(X)$ is a symmetric tensor field.   
These fluctuations are typically gapped since their shift symmetries are not supported by the underlying $\mathrm{U}(1)$ $p$-form global symmetry. 
%
%
See also Refs.~\cite{Iqbal:2021rkn,Hidaka:2023gwh} for more details.        
%

\subsection{Nonuniform bosonic surfaces}\label{trapped}
In experimental realizations of Bose-Einstein Condensation, dilute atoms are trapped by an external potential $V(x)$ induced by external magnetic fields.    
In this case, the ground state is a bound state whose characteristic quantum-mechanical length is fixed by the oscillation frequency $\omega^2=V''(0)/m$. 

Within the present framework, we can also consider analogous trapped systems. 
Here, let us focus on a class of models in which the external potential $V[C_p^{}]$ takes a separable form as 
\aln{
V[C_p^{}]=V_{\rm CM}^{}(x)+V_R^{}[\{Y\}]~,
\label{external potential}
} 
where $x$ denotes the center-of-mass position and $\{Y(\xi)\}=\{Y^k(\xi)\}_{k=1}^{D-1}$ are the relative coordinates introduced in Eq.~(\ref{COM}).   
%
%
In this case, we can employ the method of separation of variables as
\aln{
\psi_G^{}[C_p^{}]=\varphi_G^{}(x)\Psi_G^{}[\{Y\}]~
\label{separation of variables}
} %
with the normalization condition
\aln{
Q=\int [dC_p^{}]|\psi_G^{}|^2=\int d^{D-1}x|\varphi_G^{}|^2\times \int {\cal D}Yw[C_p^{}]|\Psi_G^{}|^2~. 
}
Without loss of generality, we can always adopt the normalization 
\aln{\int {\cal D}Yw[C_p^{}]|\Psi_G^{}|^2=1\quad \leftrightarrow \quad 
Q=\int d^{D-1}x|\varphi_G^{}|^2~,
}
which coincides with Eq.~(\ref{GP equation}).  
Substituting Eq.~(\ref{separation of variables}) into Eq.~(\ref{continuum surface energy}) and including the kinetic term of the center-of-mass position as well, we obtain the effective energy functional of $\varphi_G^{}(x)$:
\aln{
E[\varphi_G^{}]=\int_{\Sigma_{D-1}^{}} d^{D-1}x\left\{
\frac{1}{2m^{}}|\partial_k^{}\varphi_G^{}(x)|^2+V_{\rm CM}^{}(x)|\varphi_G^{}(x)|^2+U_{\rm eff}^{}(|\varphi_G^{}|^2)
\right\}~,
\label{effective energy functional}
}
where 
\aln{U_{\rm eff}^{}(|\varphi_G^{}|^2)=\int {\cal D}Y\omega[C_p^{}]\bigg\{|&\varphi_G^{}|^2\left(\int_{S^p}d^p\xi \frac{\sqrt{h(\xi)}}{\mathrm{Vol}[C_p^{}]}\frac{\delta \Psi_G^*}{\delta \Sigma^{k}(\xi)} \frac{\delta \Psi_G^{}}{\delta \Sigma_{k}(\xi)}+V_R^{}[\{Y\}]|\Psi_G^{}|^2
\right)
\nn
&+U\left(|\varphi_G^{}|^2|\Psi_G^{}|^2
\right)
\bigg\}~
\label{effective interaction potential}
}
is the effective potential for $\varphi_G^{}(x)$. 
The first term in the right-hand side of Eq.~(\ref{effective interaction potential}) can be absorbed into the chemical potential, while the second term leads to an effective interparticle interaction.  
As a result, Eq.~(\ref{effective energy functional}) is nothing but a conventional energy functional for interacting bosons, whose variation yields the standard Gross-Pitaevskii equation~(\ref{GP equation}).  
%
%
%

Let us study one concrete example.

\

%
\noindent{\bf HARMONIC POTENTIAL} \\
The simplest example is a spherically symmetric harmonic potential acting on the center-of-mass position:
\aln{
V[C_p^{}]=V_{\rm CM}^{}(r)=\frac{\omega_p^2}{2}r^2~,\quad r=|x|~.
}
In this case, the relative-motion part is completely irrelevant, $\psi_G^{}[C_p^{}]=\varphi_G^{}(r)$, and we can employ the Thomas-Fermi approximation~\cite{RevModPhys.71.463,RevModPhys.73.307,Pethick_Smith_2008} to examine the qualitative behavior of the ground state.    
%
Assuming the quartic interaction $U_{\rm eff}^{}(|\varphi_G^{}|^2)=\lambda (|\varphi_G^{}|^2)^2/2$, we have  
\aln{ 
&(V_{\rm CM}^{}(r)+\lambda |\varphi_G^{}(r)|^2)\varphi_G^{}(r)=\mu\varphi_G^{}(r)~,
\nn
\therefore& \quad n(r)\coloneq |\varphi_G^{}(r)|^2=\frac{\mu-V_{\rm CM}^{}(r)}{\lambda}
}
in the region where the right-hand side is positive and $\varphi_G^{}(r)=0$ otherwise. 
The boundary radius $R$ is determined by 
\aln{\mu=V_{\rm CM}^{}(r)\quad \therefore R=\frac{\sqrt{2\mu}}{\omega_p^{}}~,
} 
and the normalization condition is 
\aln{
Q=\int d^{D-1}x|\varphi_G^{}|^2=\frac{2\mu}{\lambda(D+1)}\mathrm{Vol}[B_{D-1}^{}(R)]\propto \mu^{\frac{D+1}{2}}~,
}
where $B_{D-1}^{}(R)$ is the $(D-1)$-dimensional ball with radius $R$. 
This equation provides a relation between $\mu$ and $Q$.  
The thermodynamic relation implies 
\aln{
\mu=\frac{\partial F[\varphi]}{\partial Q}=\frac{\partial E[\varphi]}{\partial Q}\bigg|_{\varphi=\varphi_G^{}}^{}~,
}
which leads to the energy per surface as 
\aln{\frac{E[\psi_G^{}]}{Q}=\frac{D+1}{D+3}\mu~.
}
In this simple trapped model, the thermodynamic properties of the ground state are completely identical to those of trapped bosons since the relative motion plays no role. 
It is particularly interesting to consider a nontrivial external potential $V_R^{}[\{Y\}]$ for the relative motion as well, and to analyze how such an internal force can modify the thermodynamic properties of the ground state. 
We leave this problem for future investigation.


\subsection{Discrete symmetries and topological order}
\label{discrete symmetry}

We can also consider discrete higher-form global symmetries.  
The simplest realization of $\mathbb{Z}_N^{[p]}$ can be obtained by adding   
\aln{
\label{ZN potential}
\lambda_N^{} \psi_G^{}[C_p^{}]^N+{\rm h.c.}~,\quad \lambda_N^{}>0~.
}
to a $\mathrm{U}(1)$ invariant potential $U(\psi_G^{*}\psi_G^{})$. 
%
Then, the continuous degeneracy of the uniform solution is lifted and reduced to a discrete set, 
\aln{
\psi_G^{}[C_p^{}]=ve^{\frac{2\pi}{N}m}~,\quad m=0,1,\cdots,N-1~.
}
Correspondingly, the phase modulation $A_p^{}$ is no longer gapless but acquires a mass gap, as we discuss below. 

In what follows, we derive the effective action of $A_p^{}$ starting from  Eq.~(\ref{non-relativistic brane theory}).  
By putting Eq.~(\ref{phase modulation}) into Eq.~(\ref{ZN potential}), one can see that $A_p^{}$ obtains a potential 
\aln{
2v^N\lambda_N^{} \cos \left(N\int_{C_p^{}}A_p^{}\right)~.
}
%
Here, we can apply the Villain formula~\cite{Villain:1977}:
\aln{
&\exp\left(-2iv^N\lambda_N^{} \int dt \int [dC_p^{}]\cos \left(N\int_{C_p^{}}A_p^{}\right)\right)
\nn
\approx &\sum_{n\in \mathbb{Z}}\exp\left(-2iv^N\lambda_N^{}\int dt\int [dC_p^{}]\left\{1-\frac{1}{2}\left(N\oint_{C_p^{}} A_p^{}-2\pi n\right)^2\right\}
\right)
}
which can be further deformed into a path-integral form by introducing two auxiliary higher-form fields as
\aln{
=\int {\cal D}f_p^{}\int {\cal D}B_{D-p-1}^{}\exp\Bigg(-&2iv^N\lambda_N^{}\int_{}dt\int [dC_p^{}]\left\{1-\frac{1}{2}\left(\oint_{C_p^{}} (NA_p^{}-f_p^{})\right)^2\right\}
\nn
&-\frac{i}{2\pi}\int_{\Sigma_D^{}}B_{D-p-1}^{}\wedge df_p^{} 
\Bigg)~,
}
where the integration of $B_{D-p-1}^{}$ imposes the flatness condition $df_p^{}=0$, which in turn corresponds to the quantization condition $\int_{C_{p}^{}}f_p^{}\in 2\pi \mathbb{Z}$. 
As shown explicitly in Ref.~\cite{Hidaka:2023gwh}, this can be further written as 
\aln{
\int {\cal D}f_p^{}\int {\cal D}B_{D-p-1}^{}\exp\left(-i\int_{\Sigma_{D}^{}}\frac{\lambda_2^{}}{2}\left(A_p^{}-\frac{f_p^{}}{N}\right)\wedge \star \left(A_p^{}-\frac{f_p^{}}{N}\right)-\frac{i}{2\pi}\int_{\Sigma_D^{}}B_{D-p-1}^{}\wedge df_p^{} 
\right)~,
} 
where $\lambda_2^{}$ is a coupling constant.  
Furthermore, $f_p^{}$ can be eliminated by using the equation of motion as
\aln{
\frac{f_p^{}}{N}=A_p^{}-\frac{N(-1)^{(D-p)(p+1)}}{2\pi \lambda_2^{}}\star dB_{D-p-1}^{}~,
} 
and we obtain the effective action:
\aln{
S_{\rm eff}^{}[A_p^{},B_{D-p-1}^{}]=(\text{kinetic terms})-
\frac{N}{2\pi}\int_{\Sigma_D^{}}
B_{D-p-1}\wedge dA_p^{}
~.
\label{BF theory}
}
In the low-energy limit, the kinetic terms are neglected, and the system is governed by the $\mathrm{BF}$-type topological field theory,  exhibiting topological order.       
In addition to the original symmetry $\mathbb{Z}_N^{[p]}$, Eq.~(\ref{BF theory}) possesses an emergent $\mathbb{Z}_N^{}$ $(D-p-1)$-form global symmetry:
\aln{
B_{D-p-1}^{}\quad \rightarrow \quad B_{D-p-1}^{}+\frac{1}{N}\Lambda_{D-p-1}^{}~,\quad d\Lambda_{D-p-1}^{}=0~,\quad \int_{C_{D-p-1}}^{}\Lambda_{D-p-1}^{}\in 2\pi \mathbb{Z}~,
\label{emergent (D-p-1) symmetry}
}
where $\Lambda_{D-p-1}^{}$ is a closed $(D-p-1)$-form and $C_{D-p-1}^{}$ is a $(D-p-1)$-dimensional closed subspace in $\Sigma_D^{}$. 
The corresponding charged object is the dual Wilson-surface operator 
\aln{
U[C_{D-p-1}^{}]=\exp\left(i\int_{C_{D-p-1}^{}} B_{D-p-1}^{}\right)~,
}
which acts on the Wilson-surface operator 
\aln{W[C_p^{}]=\exp\left(i\int_{C_p^{}} A_p^{}\right)
}
as a $\mathbb{Z}_N^{[p]}$ symmetry operator. 
%
%

\

\noindent {\bf TOPOLOGICAL ORDER}\\
Let us examine topological order in more detail.  
Here, we focus on the ground state specified by $\psi_G^{}[C_p^{}]=v$:
\aln{
|v\rangle=\sqrt{\cal N}:\exp\left(v\int[dC_p^{}]\hat{\phi}[C_p^{}]\right):|0\rangle~,
\label{Z_N states}
}
which satisfies $\hat{\phi}[C_p^{}]|v\rangle=v|v\rangle$ for all $C_p^{}\in \Gamma_p^{}$ by definition.  
The $\mathbb{Z}_N^{[p]}$ symmetry operator is 
\aln{
\hat{U}_{2\pi/N}^{}[\Sigma_{D-p-1}^{}]=\exp\left(i\frac{2\pi}{N}\int [dC_p^{}]\mathrm{I}[C_p^{},\Sigma_{D-p-1}^{}]\hat{\phi}^\dagger[C_p^{}]\hat{\phi}[C_p^{}]
\right)~.
} 
Since we focus on ground-state properties, it is convenient to express the field operator $\hat{\phi}[C_p^{}]$ effectively as 
\aln{\hat{\phi}[C_p^{}]=ve^{i\hat{\theta}[C_p^{}]}\equiv v\hat{W}[C_p^{}]~.
}
%
When the topology of spatial manifold is trivial, $\Sigma_{D-1}^{}=\mathbb{R}^{D-1}$, these operators satisfy 
\aln{
\langle v|\hat{W}[C_p^{}]|v\rangle=1~,\quad \langle v|\hat{U}_{2\pi/N}^{}[\Sigma_{D-p-1}^{}]|v\rangle=1~
}   
for all $C_p^{}$ and $\Sigma_{D-p-1}^{}$ in $\mathbb{R}^{D-1}$. 
%
Namely, all surface operators are acting on the ground state trivially.        

On the other hand, when the topology of spatial manifold is non-trivial, these operators can wind along nontrivial subspaces.   
For example, let us consider $\Sigma_{D-1}^{}=S^p\times S^{D-p-1}$, namely, $C_p^{}=S^p$ and $\Sigma_{D-p-1}^{}=S^{D-p-1}$. 
In this case, $\hat{W}[S^{p}]$ and $\hat{U}_{2\pi/N}^{}[S^{D-p-1}]$ are  linked, i.e.,  
\aln{
\hat{U}_{2\pi/N}^{}[S^{D-p-1}]\hat{W}[S^{p}]\hat{U}_{2\pi/N}^{-1}[S^{D-p-1}]=e^{i\frac{2\pi}{N}}\hat{W}[S^{p}]~
\label{operator relation}
}
as an (equal-time) operator relation. 
Then, consider a new state
\aln{
|v'\rangle\coloneq \hat{U}_{2\pi /N}^{}[S^{D-p-1}]^{-1}|v\rangle~,
}
which has the same energy as $|v\rangle$ because $\hat{U}_{2\pi/N}^{}[S^{D-p-1}]$ is the symmetry operator. 
Using Eq.~(\ref{operator relation}), one can see
\aln{
\hat{W}[S^{p}]|v'\rangle&=\hat{U}[S^{D-p-1}]^{-1}\bigg(\hat{U}[S^{D-p-1}]^{}\hat{W}[S^p]\hat{U}[S^{D-p-1}]^{-1}\bigg)|v\rangle
\nn
&=e^{i\frac{2\pi}{N}}|v'\rangle~,
}
namely, $|v'\rangle$ is another eigenstate of $\hat{W}[S^p]$ with the same energy as $|v\rangle$.   
By repeating the same procedure, one can see that there are $N$ degenerate ground states 
\aln{
\left(\hat{U}_{2\pi /N}[S^{D-p-1}]^{-1}\right)^m|v\rangle\quad m=0,1,\cdots,N-1~,
} 
when $\Sigma_{D-1}^{}=S^p\times S^{D-p-1}$. 
In general, the ground-state degeneracy depends on the dimensions $D,p$, as well as on the topology of $\Sigma_{D-1}^{}$. 
For instance, when $D=3$, $p=1$, and $\Sigma_{2}^{}=T^2$, each loop  operator can wind independently along either of two non-contractible loops of $T^2$. 
Consequently, the degeneracy becomes $N^2$.

\

\noindent {\bf SURFACE EXCITATIONS}\\
%
\begin{figure}
    \centering
     \includegraphics[scale=0.6]{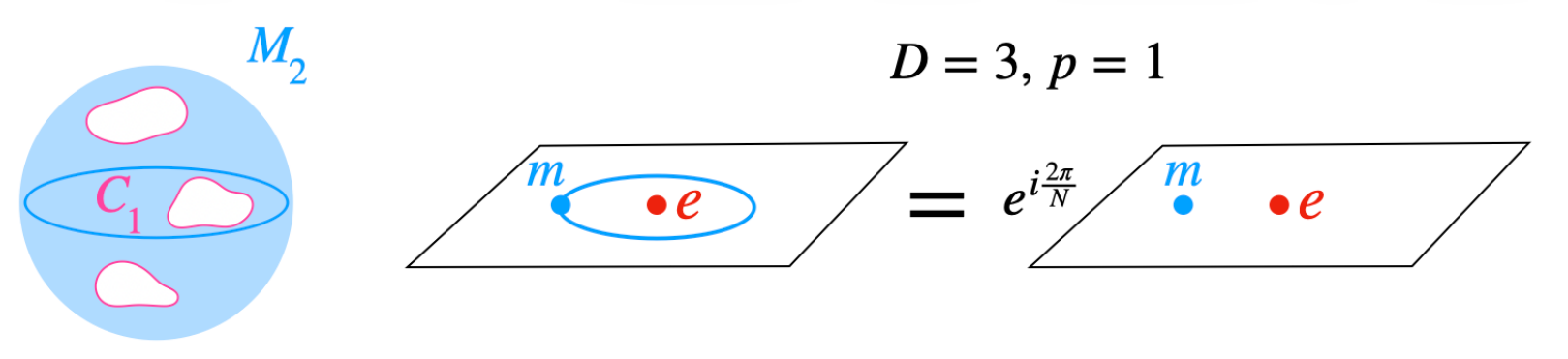}
    \caption{
   Left: Three loop excitations for $p=2$.   
    Right: Braiding phase when $D=3$ and $p=1$. 
    }
    \label{fig:anyon}
\end{figure}
To discuss excitations in the condensed phase, we return to the hypercubic spatial lattice $\Lambda_{D-1}^{}$ instead of a continuous spatial manifold.    
Up to this point, we have only considered the closed surface operators $\hat{\phi}[C_p^{}]$, $\hat{U}_{2\pi/N}^{}[\Sigma_{D-p-1}^{}]$.  
However, it is also possible to define {\it open} surface operators     
\aln{
\hat{W}[M_p^{}]&=\frac{1}{v}\prod_{\sigma_p^{}\in M_p^{}}\hat{\phi}(\sigma_p^{})~,\quad \partial  M_{p}^{}=C_{p-1}^{}\neq \emptyset~,
\\
\hat{U}_{2\pi/N}^{}[M_{D-p-1}^{}]&=e^{i\frac{2\pi}{N}
\hat{Q}_p^{}[M_{D-p-1}^{}]}~,\quad \partial  M_{D-p-1}^{}=C_{D-p-2}^{}\neq \emptyset~,
}
where $M_p^{}~(M_{D-p-1}^{})$ is a connected $p~(D-p-2)$-dimensional surface bounded by a $(p-1)~(D-p-2)$-dimensional closed surface $C_{p-1}^{}~(C_{D-p-2}^{})$, which can contain multiple disconnected regions, as illustrated in the left panel of Fig.~\ref{fig:anyon}. 
In this example, $M_2^{}$ has three $S^1$ boundaries. 
The topological nature of the original closed surface operators implies that the energies of the states 
\aln{
|e\rangle_{p-1}^{}\coloneq \hat{W}[M_p^{}]|v\rangle~,\quad |m\rangle_{D-p-2}^{}\coloneq \hat{U}_{2\pi/N}^{}[M_{D-p-1}^{}]|v\rangle~
}
depend only on their boundary manifolds and thus represent   
%
%
boundary excitations.    
For example, when $C_{p-1}^{}$ consists of $n$ disconnected components as 
\aln{
C_{p-1}^{}=\cup_{i=1}^n C_{p-1}^{(i)}~,\quad C_p^{(i)}\cap C_p^{(j)}=\emptyset \quad \text{for } i\neq j~,
}
$|e\rangle_{p-1}^{}$ represents an excitation of $n$ $(p-1)$-dimensional closed surfaces, as shown in the left panel in Fig.~\ref{fig:anyon} for $p=2$ and $n=3$.  
In particular, the boundaryless limit $C_{p-1}^{}\rightarrow \emptyset$ corresponds to the annihilation of these surfaces, resulting in the trivial state, $\lim_{C_{p-1}^{}\rightarrow \emptyset}|e\rangle_{p-1}^{}=|v\rangle$, by definition. 
%
%

Moreover, the commutation relation implies 
\aln{
\hat{U}_{2\pi/N}^{}[\Sigma_{D-p-1}^{}]|e\rangle_{p-1}^{}&=\hat{U}_{2\pi/N}^{}[\Sigma_{D-p-1}^{}]\hat{W}[M_{p}^{}]\hat{U}[\Sigma_{D-p-1}^{}]^{-1}|v\rangle
\nn
&=e^{i\frac{2\pi}{N}\mathrm{I}[M_{p}^{},\Sigma_{D-p-1}^{}]}|e\rangle_{p-1}^{}~,
}
which indicates the presence of a fractional braiding phase when $|e\rangle_{p-1}^{}$ and $|m\rangle_{D-p-2}^{}$ are linked each other.  
In the right panel of Fig.~\ref{fig:anyon}, we illustrate the case for $D=3$ and $p=1$, in which both excitations are particle-like and are known as (abelian) {\it anyons}. 
%




\subsection{Topological defects}

In conventional bosonic systems such as Eq.~(\ref{particle Hamiltonian}), the Gross-Pitaevskii equation~(\ref{GP equation}) admits topologically nontrivial (static) solutions, namely, topological defects in general.
Here, we demonstrate that our variational equation~(\ref{continuum surface variational equation}) similarly admits topologically nontrivial solutions. 
For simplicity, we consider the flat space $\Sigma_{D-1}^{}=\mathbb{R}^{D-1}$ below. 

\

\noindent {\bf $\mathrm{U}(1)^{[p]}$ CASE}\\
%
\begin{figure}
    \centering
    \includegraphics[scale=0.2]{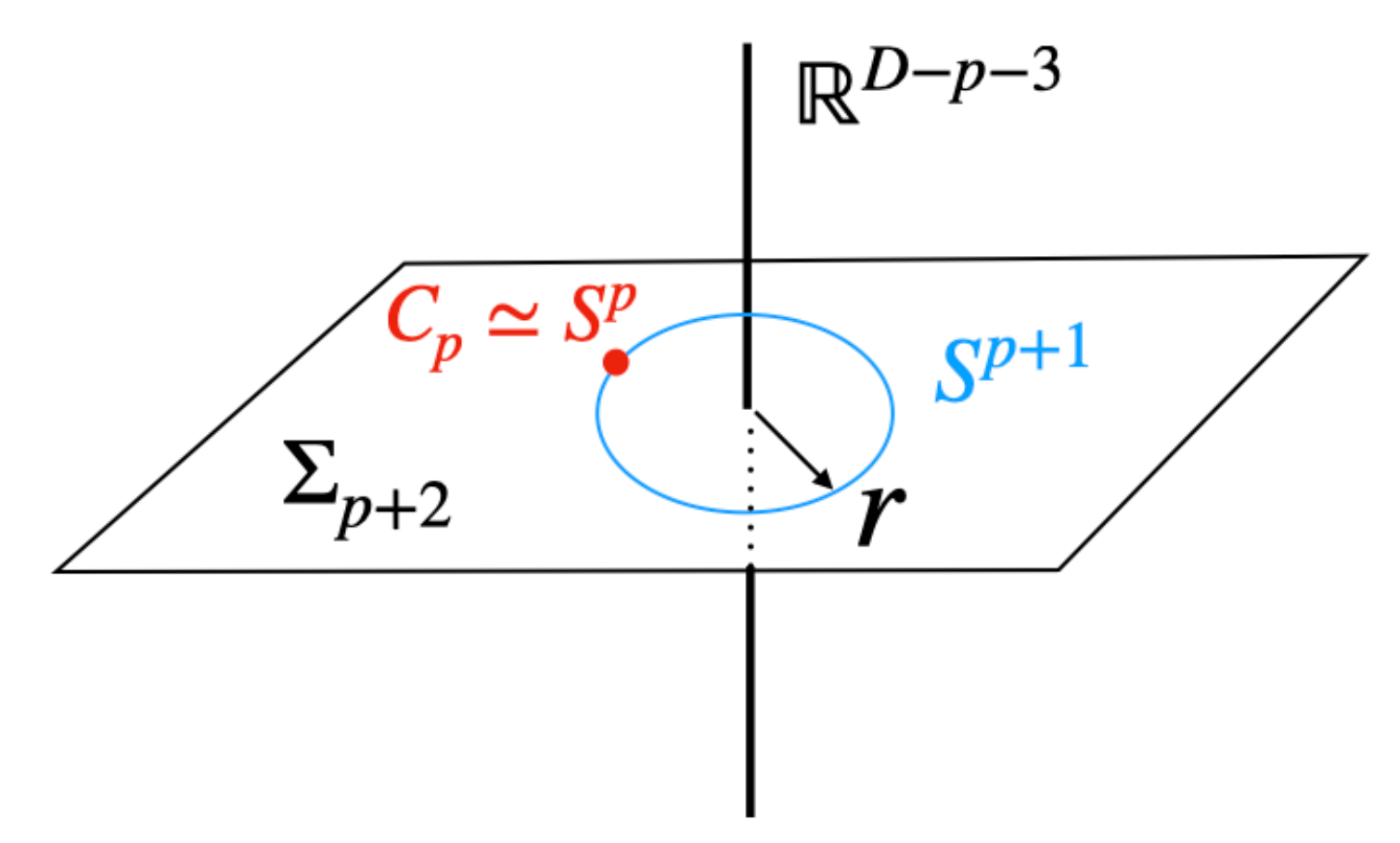}
    \caption{
   A higher-dimensional vortex solution for $G^{[p]}=\mathrm{U}(1)^{[p]}$.     
    The red point corresponds to the embedded $p$-brane $C_p^{}\simeq S^p$ winding along $S^{p+1}$.  
    }
    \label{fig:U(1)vortex}
\end{figure}
Let us first consider $G^{[p]}=\mathrm{U}(1)^{[p]}$ and represent $\mathbb{R}^{D-1}$ as 
\aln{
\mathbb{R}^{D-1}=\mathbb{R}^{D-p-3}\times \Sigma_{p+2}^{}\coloneq \mathbb{R}^{D-p-3}\times S^{p+1}\times [0,\infty)~,
}
where the coordinate $r$ of $[0,\infty)$ corresponds to the radius of the $(p+2)$-dimensional subspace $\Sigma_{p+2}^{}$. 
See Fig.~\ref{fig:U(1)vortex} as an illustration.  
Besides, we represent the volume form of the $q$-dimensional sphere $S^q$ as $\Omega_q^{}$, i.e.,
\aln{
\int_{S^q} \Omega_q^{}=\mathrm{Vol}[S^{q}]~.
}
For $q=0$, we define $\mathrm{Vol}[S^{0}]=1$.

We consider a spherically symmetric $p$-brane configuration $C_p^{}\simeq S^p$ embedded within $S^{p+1}$ and parameterized by $(r,\theta_1^{})$, where $\theta_1^{}$ is one of the polar coordinates $\theta_1^{}\in [0,\pi]$ of $S^{p+1}$.  
This is schematically shown by a red point in Fig.~\ref{fig:U(1)vortex}. 
This spherical configuration corresponds to reducing the (path-)integral measure of $C_p^{}$ as
\aln{
\int [dC_p^{}]\quad \rightarrow \quad 
\int_0^\infty dr r^{p+1}\int_0^{\pi} d\theta_1^{}(\sin\theta_1^{})^p
~. 
\label{minispace}
}  
Besides, the volume of $C_p^{}$ is explicitly given by  
\aln{
{\rm Vol}[C_p^{}]=(r\sin\theta_1^{})^p\int_{S^{p}}\Omega_p^{}
=(r\sin\theta_1^{})^p\times \mathrm{Vol}[S^{p}]~.
}
%
%
We then consider the following ansatz
\aln{
\psi_{}^{}[C_p^{}]=\frac{1}{\sqrt{2}}\left(\int_{C_{p}^{}}\chi_p^{}\right)
\exp\left(i\int_{C_p^{}}A_{p}^{\text{defect}}\right)~,
\label{U(1) ansatz}
}
where
\aln{
\chi_p^{}
=\chi(r)(\sin\theta_1^{})^{p}\Omega_p^{}~,\quad \chi(r)\in \mathbb{R}~,
}
and $F_{p+1}^{\text{defect}}:=dA_{p}^{\text{defect}}$ is the normalized volume-form on $S^{p+1}=C_p^{}\times [0,\pi]$ satisfying 
\aln{
\int_{S^{p+1}} F_{p+1}^{\text{defect}}=2\pi q~,\quad q\in \mathbb{Z}~.
\label{theta normalization}
}   
This normalization guarantees the single-valuedness of the wave function $\psi[C_p^{}]$ in the volumeless limit as
%
\aln{\psi[C_p^{}]|_{\theta_1^{}=0}=\psi[C_p^{}]|_{\theta_1^{}=\pi}~.
}
%
Note that Eq.~(\ref{theta normalization}) is nothing but the Dirac quantization condition for the field strength of $A_p^{}$ and is regarded as a higher-form version of the quantization of circulation
\aln{\oint_{S^1} d \mathbf{l}\cdot \nabla \phi\in 2\pi \mathbb{Z}~.
}

Denoting the world-manifold of the defect as 
\aln{
M^{}_{D-p-2}\coloneq \mathbb{R}\times \mathbb{R}^{D-p-3}~,
}
we can also relate $F_{p+1}^{\text{defect}}$ to the Poincare dual-form of the world-manifold:\footnote{
This expression of the Poincare dual is valid only for spherically symmetric  functions. 
Otherwise, we must include the delta functions for angular variables.     
}
\aln{\delta_{p+2}^{}(M^{}_{D-p-2})=\frac{1}{2\pi q}dF_{p+1}^{\text{defect}}~.
\label{Poincare dual and magnetic field}
}
%
In fact, one can check 
\aln{
\int_{\Sigma_{D}^{}}\delta_{p+2}^{}(M^{}_{D-p-2})\wedge \delta_{D-p-2}^{}(\Sigma_{p+2}^{})&=\int_{\Sigma_{p+2}}\delta_{p+2}^{}(M^{}_{D-p-2})=1~,
}
which represents the linking between $M_{D-p-2}^{}$ and $S_{}^{p+1}$. 
%
\begin{figure}
    \centering
    \includegraphics[scale=0.5]{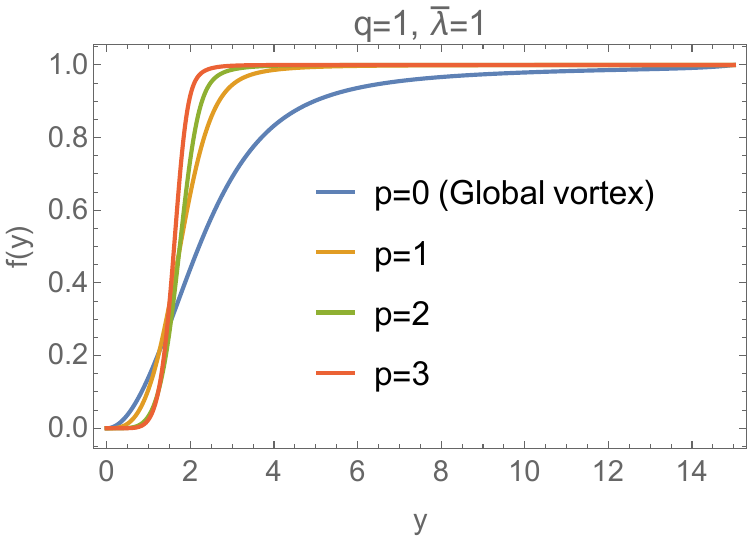}
    \caption{
    Field profiles of the topological defect for $G=\mathrm{U}(1)$. 
    Each color corresponds to a different choice of $p$.  
    }
    \label{fig:field profile for U(1)}
\end{figure}

%
The area derivative on the ansatzs~(\ref{U(1) ansatz}) is evaluated as 
\aln{
\frac{\delta \phi[C_p^{}]}{\delta \Sigma^{\mu_1^{}\cdots \mu_{p+1}}(\xi)}&=\frac{1}{\sqrt{2}}\left(d\chi_p^{}+i\left(\int_{C_p^{}}\chi_p^{}\right)dA_p^{\text{defect}}\right)_{\mu_1^{}\cdots \mu_{p+1}^{}}^{}e^{i\int_{C_p^{}}A_p^{\text{defect}}}
~,
}
by which we can evaluate the effective energy of $\chi(r)$ as
\aln{
&\int_0^{\infty}dr\int_0^{\pi}d\theta_1^{}\frac{(\sin\theta_1^{})^p}{r^{p-1}}\left[\frac{1}{2}\left(\frac{d\chi}{dr}\right)^2+\frac{1}{2r^2}\left\{\frac{p^2}{(\sin\theta_1^{})^2}+\left(\frac{2\pi q\mathrm{Vol}[S^{p}](\sin\theta_1^{})^p}{\mathrm{Vol}[S^{p+1}]}\right)^2\right\}\chi^2+r^{2p}U(\phi)
\right]
\nn
&\propto \int_0^{\infty}dr\frac{1}{r^{p-1}}\left[\frac{1}{2}\left(\frac{d\chi}{dr}\right)^2+\frac{c_2^{}}{2r^2}\chi^2+r^{2p} \tilde{U}(\chi)
\right]~,
\label{vortex action}
}
where $c_2^{}$ is a numerical constant determined by $p$ and $q$, and $\tilde{U}(\chi)$ is the effective potential of $\chi$ after integrating out $\theta_1^{}$. 
This equation reduces to the conventional vortex energy for $p=0$. 
By introducing dimensionless quantities
\aln{
f=\frac{\chi}{v}~,\quad y=vr~,\quad \overline{U}(f)=\frac{1}{v^4}\tilde{U}(\chi)~, 
} 
we obtain the dimensionless equation of motion: 
\aln{
\frac{1}{y^{p+1}}\frac{d}{dy}\left(\frac{1}{y^{p-1}}\frac{df}{dy}\right)-\frac{c_2^{}}{y^{2(p+1)}}f-\frac{\delta \overline{U}(f)}{\delta f}=0~,
\label{dimensionless eom}
}
where $f(y)$ satisfies the boundary condition: $f(0)=0$ and $f(\infty)=1$. 
For simplicity, we consider the $\mathrm{U}(1)$ invariant quartic potential 
\aln{\overline{U}(f)=\frac{\overline{\lambda}}{4}(f^2-1)^2~,\quad \lambda>0~,
}
as an example.   
In Fig.~\ref{fig:field profile for U(1)}, we show the numerical plots of  $f(y)$ for $p=0$ (blue), $p=1$ (orange), $p=2$ (green), and $p=3$ (red) with $\overline{\lambda}=1$ and $q=1$. 
The blue one corresponds to the conventional global vortex profile.  
One can see that the variation in $p$ corresponds to the change of the slope of $f(y)$ in the vicinity of the origin, as expected from the equation of motion~(\ref{dimensionless eom}).   

\

\noindent {\bf $\mathbb{Z}_N^{[p]}$ CASE}\\
Next let us consider the discrete higher-form symmetry $G^{[p]}=\mathbb{Z}_N^{[p]}$. 
%
%
In this case, we represent $\mathbb{R}^{D-1}$ as 
\aln{
\mathbb{R}^{D-1}=\mathbb{R}^{D-p-2}\times \Sigma_{p+1}^{}\coloneq \mathbb{R}^{D-p-2}\times S^{p}\times [0,\infty)~,
\label{manifold decomposition}
}
where $r\in [0,\infty)$ denotes the radius of $S^p$.  
%
%
Note that the boundary of $\Sigma_{p+1}^{}$ is  $\partial \Sigma_{p+1}^{}= \Sigma_{p+1}^{}|_{r=\infty}^{}\simeq S^{p}$.  
%
\begin{figure}
    \centering
    \includegraphics[scale=0.3]{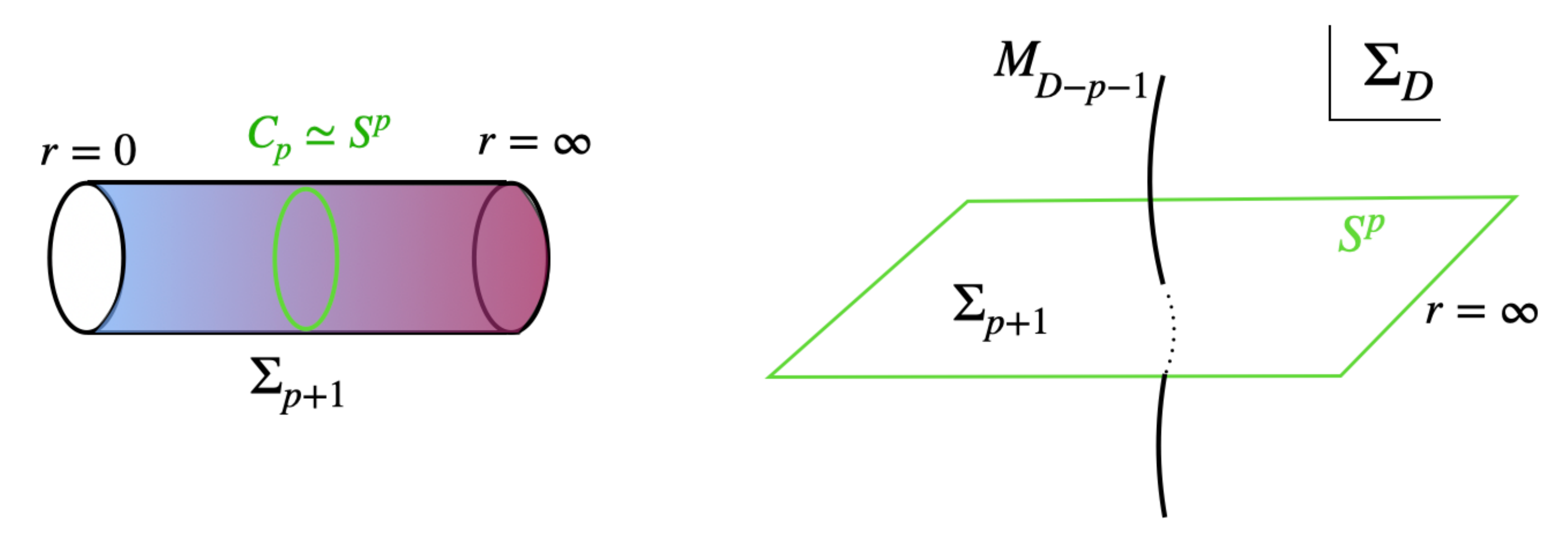}
    \caption{
    Left: Topologically nontrivial static configuration for $G^{[p]}=\mathbb{Z}_N^{[p]}$.     
    The green circle represents an embedded surface $C_p^{}\simeq S^p$.   
    \\
    Right: Linking between the world manifold $M^{}_{D-p-1}$ of topological defect and $S^{p}$. 
    }
    \label{fig:soliton}
\end{figure}
%
We then consider a surface $C_p^{}$ embedded into $S^p$ in Eq.~(\ref{manifold decomposition}), as depicted by a green circle in the left panel in Fig.~\ref{fig:soliton}.  
This corresponds to the reduction of the path-integral measure as 
\aln{
\int [dC_p^{}] \quad \rightarrow \quad \int_0^\infty dr r^{p}
~.
\label{minispace}
} 
We then consider the following ansatz:
\aln{
\psi[C_p^{}]=\frac{v}{\sqrt{2}}\exp\left(i\int_{S^p}A_p^{\text{defect}}\right)~,
\quad A_{p}^{\text{defect}}=f(r)\frac{\Omega_{p}^{}}{\mathrm{Vol}[S^p]}~,
\label{ansatz}
}
with the boundary conditions
\aln{
f(0)=0~,\quad f(\infty)=\frac{2\pi}{N}n~,\quad n\in \mathbb{Z}~. 
}
Namely, Eq.~(\ref{ansatz}) represents a topological configuration interpolating two degenerate minima.  
The corresponding topological charge is 
\aln{
Q_{D-p-1}^{}=\frac{N}{2\pi}\int_{\Sigma_{p+1}^{}}dA_p^{\text{defect}}=\frac{N}{2\pi}\int_{S^p}A_p^{\text{defect}}\bigg|_{r=\infty}^{}=n\in \mathbb{Z}~,
}
where we have used the Stokes theorem in the second equality.

Denoting the world manifold of the defect as 
\aln{
M^{}_{D-p-1}\coloneq \mathbb{R}\times \mathbb{R}^{D-p-2}~,
}
we can relate $A_p^{\text{defect}}$ to the Poincare dual of $M^{}_{D-p-1}$ as  
\aln{
\delta_{p+1}^{}(M^{}_{D-p-1})=\frac{N}{2\pi Q_{D-p-1}^{}}dA_p^{\text{defect}}~.\label{world-volume form}
}
In fact, we can check 
\aln{
\int_{\Sigma_{D}^{}}\delta_{p+1}(M^{}_{D-p-1})\wedge \delta_{D-p-1}^{}(\Sigma_{p+1}^{})=\int_{\Sigma_{p+1}}\delta_{p+1}^{}(M^{}_{D-p-1})=\frac{N}{2\pi Q_{D-p-1}^{}}\int_{\Sigma_{p+1}}dA^{\text{defect}}_{p}=1~,
}
which represents the linking between $M_{D-p-1}^{}$ and $C_p^{}\simeq S_{}^{p}$, as illustrated in the right panel in Fig.~\ref{fig:soliton}.  

The area derivative on the ansatz~(\ref{ansatz}) can be evaluated as 
\aln{
\frac{\delta \psi[C_p^{}]}{\delta \Sigma^{\mu_1^{}\cdots \mu_{p+1}^{}}(\xi)}=
i (dA_p^{\text{defect}})_{\mu_1^{}\cdots \nu_{p+1}^{}}\psi[C_p^{}]=-i\frac{1}{\mathrm{Vol}[S^{p}]r^p}\frac{df}{dr}\delta_\mu^{r}\psi[C_p^{}]~,
}
which leads to the following effective energy of $f(r)$:
\aln{
\int_0^\infty drr^p\left[\frac{v^2}{2\mathrm{Vol}[S^p]^2}\left(\frac{1}{r^p}\frac{df}{dr}\right)^2+U(f)\right]
\propto v^{p+3}\int_0^\infty dyy^p\left[\frac{1}{2}\left(\frac{1}{y^p}\frac{df}{dy}\right)^2+\overline{U}(f)\right]~,
\label{defect effective action}
}
where $y\coloneq (\mathrm{Vol}[S^p])^{\frac{1}{p+1}}vr$ and $\overline{U}\coloneqq \mathrm{Vol}[S^p]^2 U/v^{2p+4}$ are the dimensional radius and potential respectively. 
By taking the variation in Eq.~(\ref{defect effective action}) with respect to $f(y)$, we obtain the dimensionless equation of motion:
\aln{
\frac{1}{y^p}\frac{d}{dy}\left(\frac{1}{y^p}\frac{df}{dy}\right)-\frac{\delta \overline{U}}{\delta f}=0~.
\label{defect equation}
}
In general, the periodic potential $\overline{U}(f)=\overline{U}\left(f+\frac{2\pi}{N}\right)$ is given by a Fourier series of $\cos(N k f)$ and $\sin(Nk f)$ with $k\in \mathbb{N}$.  
For a concrete example, let us consider the simplest one   
\aln{
\overline{U}(f)=\overline{U}_A^{}(f)\coloneq \overline{\lambda}\left(1-\cos\left(N f\right)\right)~,
\label{cosine potential}
} 
where $\overline{\lambda}$ is a dimensionless coupling. 
%
%
In this case, $N$ dependence is essentially irrelevant because it can be eliminated in the equation of motion~(\ref{defect equation}) by  $Nf\rightarrow f$ and $N^{\frac{1}{p}}y\rightarrow y$.  
Thus, it suffices to calculate the profile $f(y)$ for $N=1$. 
\begin{figure}
    \centering
    \includegraphics[scale=0.65]{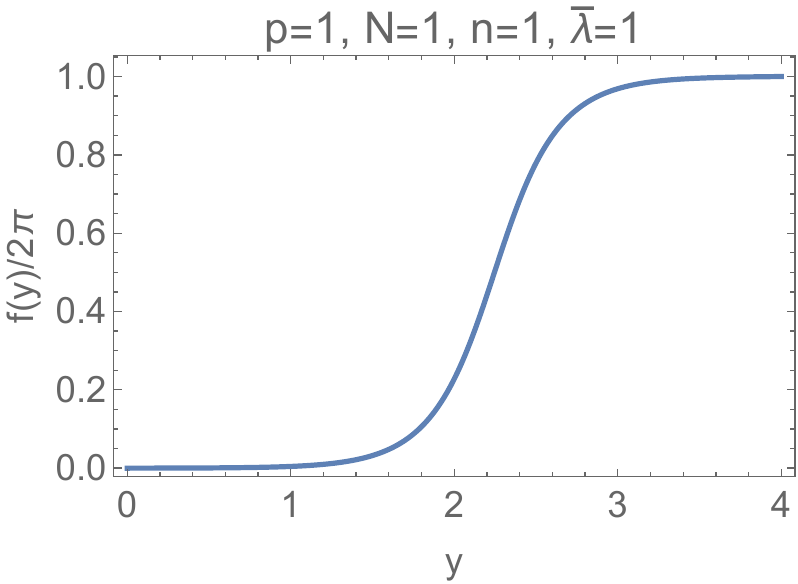}
    \caption{
    Field profile of the topological defect for $G^{[p]}=\mathbb{Z}_N^{[p]}$.  
    }
    \label{fig:profile}
\end{figure}
\noindent 
In Fig.~\ref{fig:profile}, we show the numerical plot of $f(y)/2\pi$ for $N=1$, $p=1$, $n=1$ and $\overline{\lambda}=1$.       

We can also obtain the low-energy effective theory in the presence of    topological defect.  
Instead of Eq.~(\ref{phase modulation}), the phase modulation $A_p^{}$ is introduced in the defect background as 
\aln{
\phi[C_p^{}]=\frac{v}{\sqrt{2}}\exp\left(i\int_{C_p^{}} (A_p^{\text{defect}}+A_p^{})\right)~, 
} 
and the low-energy effective action is obtained by replacing $A_p^{}$ in Eq.~(\ref{BF theory}) by $A_p^{}+A_p^{\text{defect}}$:
\aln{
S_{\rm eff}^{}[A_p^{},B_{D-p-1}^{}]\approx \frac{N}{2\pi}\int_{\Sigma_{D}^{}}B_{D-p-1}^{}\wedge dA_p^{}+Q_{D-p-1}^{}\int_{M_{D-p-1}^{}}B_{D-p-1}^{}~,
\label{effective theory in discrete}
}
where we have used Eq.~(\ref{world-volume form}). 
The second term is an interaction between the dual field $B_{D-p-1}^{}$ and the topological defect as expected.  
%


\section{$\mathbb{Z}_N^{}$ $p$-form lattice gauge theory}\label{sec:model}

In this section, we apply our variational method to a $\mathbb{Z}_N^{}$ $p$-form lattice gauge theory.     
Let us summarize our notation first.    

In general, we consider a $(D-1)$-dimensional spatial complex $\Delta_{D-1}^{}$ as $\Sigma_{D-1}^{}$, whose topology is of interest and is composed of simply-connected cells.
Here, ``simply-connected" means that each cell is topologically a ball of the corresponding dimension.  
The set of all unorientated $p$-dimensional cells are denoted by $\Delta_p^{}$, and a generic element is written as $\sigma_p^{}\in \Delta_p^{}$.  
We refer to $\sigma_p^{}$ simply as $p$-cell.  
For each cell, an orientation can be assigned, which we distinguish by $\pm \sigma_p^{}$.\footnote{More concretely, when a $p$-cell $\sigma_p^{}$ consists of $m$ sites labeled by $m$ numbers $i_1^{}<i_2^{}<\cdots<i_m^{}$, we define a positively (negatively) oriented cell by $(\pm)\sigma_p^{}=\langle i_{f(1)}^{}i_{f(2)}^{}\cdots i_{f(m)}^{}\rangle$ for even (odd) permutations $f\in S_m^{}$.  
}   
Moreover, a general $p$-dimensional closed surface $C_p^{}$ is constructed by a finite number of $p$-cells, whose total number is denoted as $|C_p^{}|$. 
For example, when $\Delta_{D-1}^{}$ is a simple hypercubic lattice, $\sigma_p^{}$ is a $p$-dimensional hypercube, and $|\partial \sigma_{p+1}^{}|=2(p+1)$.

The $p$-chain group is defined by 
\aln{
\Omega_p^{}\coloneq \left\{\sum_{\sigma_p^{}\in \Delta_p^{}}c_i^{}\sigma_p^{}|c_i^{}\in \mathbb{Z}_2^{}\right\}~,
}
The boundary operator $\partial:\Omega_p^{}~\rightarrow~\Omega_{p-1}^{}$ is introduced by a linear operator which maps $\sigma_p^{}$ into a linear combination of $\sigma_{p-1}^{}\in \partial \sigma_{p}^{}$ along with the nilpotency condition $\partial^2=0$. 
For example, when $\Delta_{D-1}^{}$ is a simplicial complex, $\sigma_p^{}$ contains $p+1$ vertices labeled by $i_0^{}<i_1^{}<\cdots <i_{p}^{}$ and is represented as $\sigma_p^{}=(i_0^{}i_1^{}\cdots i_{p}^{})$.  
Then, the boundary operator is defined by 
\aln{
\partial\sigma_p^{}=\sum_{l=0}^{p}l(i_0^{}\cdots i_{k-1}^{}\hat{i}_{k}^{}i_{k+1}^{}\cdots i_{p}^{})~.
}
For a general complex $\Delta_{D-1}^{}$, we can define $\partial$ in a similar manner.  
%
%

Besides, we can naturally introduce an inner product in each $\Omega_p^{}$ by 
\aln{
\left\langle \sigma_p^{}~,\sigma_p^{'}
\right\rangle=\begin{cases}1  & \text{for }\sigma_p^{}=\sigma_p^{'}
\\
0  & \text{for }\sigma_p^{}\neq \sigma_p^{'}
\end{cases}~.
}
By using this, the adjoint operator $\overline{\partial}:\Omega_p^{}\rightarrow \Omega_{p+1}^{}$ is defined by 
\aln{\overline{\partial}\sigma_p^{}=\sum_{\substack{\sigma_{p+1}^{}\in \Delta_{p+1}^{}~{\rm s.t.}\\ \langle \partial \sigma_{p+1}^{},\sigma_p^{}\rangle=k}}k\sigma_{p+1}^{}~.
}
By definition, it satisfies 
\aln{\label{adjoint relation}
\langle \sigma_{p+1}^{},\overline{\partial}\sigma_p^{}\rangle=\langle \partial \sigma_{p+1}^{},\sigma_p^{}\rangle\quad \text{for }  {}^\forall \sigma_{p+1}^{}\in \Delta_{p+1}^{}~\text{and}~ {}^\forall \sigma_{p}^{}\in \Delta_{p}^{}~. 
}
%
Furthermore, the nilpotency property of $\partial$ and $\overline{\partial}$ implies  
\aln{
\sum_{\sigma_p^{}\in \Delta_p^{}}\langle\partial\sigma_{p+1}^{},\sigma_p^{}\rangle \langle \partial \sigma_p^{}, \sigma_{p-1}^{}\rangle=0~,
\label{nilpotency1}
\\
\sum_{\sigma_p^{}\in \Delta_p^{}}\langle\sigma_{p+1}^{},\overline{\partial}\sigma_p^{}\rangle \langle \sigma_p^{}, \overline{\partial}\sigma_{p-1}^{}\rangle=0~,
\label{nilpotency2}
}
for arbitrary $\sigma_{p-1}^{}$ and $\sigma_{p+1}^{}$.\footnote{
There equations are matrix expressions of $\partial^2=0$ and $\overline{\partial}^2=0$ respectively.    
}

Finally, any quantity or operator defined on the dual complex will be denoted with a caron, such as $\check{A}$.   
In particular, we denote the dual $(D-p-1)$-dimensional cell to $\sigma_p^{}$ by $\check{\sigma}_{D-p-1}^{}$. 
For instance, when $D=3$ and $p=1$, we have $\check{\sigma}_{D-p-1}^{}=\check{\sigma}_1^{}$, which is a link.  
For $D=4$ and $p=1$, we have $\check{\sigma}_{D-p-1}^{}=\check{\sigma}_2^{}$, which is a face. 

\subsection{Hamiltonian and symmetry}\label{Higher gauge theory on the lattice}

We assign a $N$-dimensional Hilbert space ${\cal H}_N^{}\coloneq \text{span}\{|n\rangle,n=0,1,\cdots,N-1\}$ on each $(p-1)$- and $p$-cell.  
Analogously to $\hat{Z}$ and $\hat{X}$ operators in qubits, we define 
\aln{
\hat{Z}_N^{}\coloneq \sum_{n=0}^{N-1} \omega^n|n\rangle \langle n|~,\quad \hat{X}_N^{}\coloneq \sum_{n=0}^{N-1} |n\rangle \langle n+1|~,
} 
where $\omega=e^{\frac{2\pi i}{N}}$. 
One can check the commutation relation $\hat{X}_N^{}\hat{Z}_N^{}=\omega \hat{Z}_N^{}\hat{X}_N^{}$. 
Apparently, the inverse operators are 
\aln{
\hat{Z}_N^{-1}=\hat{Z}_N^{\dagger}=\sum_{n=0}^{N-1} \omega^{-n}|n\rangle \langle n|~,\quad \hat{X}_N^{-1}=\hat{X}_N^{\dagger}=\sum_{n=0}^{N-1} |n+1\rangle \langle n|~. 
}
When these operators are defined on a $p$-cell $\sigma_p^{}$, we denote them as $\hat{Z}_N^{}(\sigma_p^{}),\hat{X}_N^{}(\sigma_p^{})$. 
For the oppositely oriented $p$-cell $-\sigma_p^{}$, we define $\hat{Z}_N^{}(-\sigma_p^{})=\hat{Z}_N^{-1}(\sigma_p^{}),\hat{X}_N^{}(-\sigma_p^{})=\hat{X}_N^{-1}(\sigma_p^{})$.
Moreover, the eigenstates of $\hat{X}_N^{}$ are denoted by $|n\rangle_X^{}~(n=0,1,2,\cdots,N-1)$.

A Hamiltonian of $\mathbb{Z}_N^{}$ $p$-form gauge theory is   
\aln{
\hat{H}=-g\sum_{\sigma_{p+1}^{}\in \Delta_{p+1}^{}}\hat{B}(\sigma_{p+1}^{})-\kappa \sum_{\sigma_p^{}\in \Delta_p^{}}\hat{B}(\sigma_p^{})\hat{Z}_N^{}(\sigma_p^{})
-h\sum_{\sigma_p^{}\in \Delta_p^{}}\hat{X}_N^{}(\sigma_p^{})-g'\sum_{\sigma_{p-1}^{}\in \Delta_{p-1}^{}}\hat{X}_N^{}(\sigma_{p-1}^{})+{\rm h.c.}~,
\label{ZN gauge theory}
}
where 
\aln{
\hat{B}(\sigma_{p+1}^{})\coloneq \prod_{\sigma_p^{}\in \Delta_p^{}
}^{}\hat{Z}_N^{\langle \partial\sigma_{p+1}^{},\sigma_p^{}\rangle}(\sigma_{p}^{})
}
is the Wilson surface operator.  
Here, we define $\hat{Z}_N^{0}=1$. 
The last two terms in Eq.~(\ref{ZN gauge theory}) correspond to transverse  magnetic interactions in Ising models.  
Assuming that all coupling constants are real, $g,h$ and $g'$ can be taken to be positive without loss of generality. 

Equation~(\ref{ZN gauge theory}) is invariant under the following $\mathbb{Z}_N^{}$ $(p-1)$-form gauge transformation:
\aln{
&\hat{Z}_N^{}(\sigma_{p-1}^{})\quad \rightarrow \quad \omega^{l(\sigma_{p-1}^{})} \hat{Z}_{N}^{}(\sigma_{p-1}^{})~,\quad l(\sigma_{p-1}^{})\in \{0,1,\cdots,N-1\}~,
\\
&\hat{Z}_{N}^{}(\sigma_p^{})\quad \rightarrow \quad \left(\prod_{\sigma_{p-1}^{}\in \Delta_{p-1}^{}
}\omega^{-\langle \partial \sigma_p^{},\sigma_{p-1}^{}\rangle l(\sigma_{p-1}^{})}\right)\hat{Z}_N^{}(\sigma_{p}^{})~,
\\
&\hat{X}_{N}^{}(\sigma_{p-1}^{})\quad \rightarrow \quad \hat{X}_{N}^{}(\sigma_{p-1}^{})~,\quad \hat{X}_{N}^{}(\sigma_{p}^{})\quad \rightarrow \quad\hat{X}_{N}^{}(\sigma_{p}^{})~,
} 
which is generated by the following local operator 
\aln{
\hat{G}(\sigma_{p-1}^{})\coloneq \hat{X}_{N}(\sigma_{p-1}^{})
\left(\prod_{\sigma_{p}^{}\in \Delta_p^{}
}\hat{X}_{N}^{\langle \sigma_p^{},\overline{\partial}\sigma_{p-1}^{}\rangle}(\sigma_p^{})\right)^{-1}\equiv  \hat{X}_{N}(\sigma_{p-1}^{})\hat{A}^{}(\sigma_{p-1}^{})^{-1}~.
\label{gauge operator}
}
%
Here, $\hat{A}(\sigma_{p-1}^{})$ can be regarded as a dual $(D-p-1)$-dimensional Wilson operator by interpreting that $\hat{X}_N^{}(\sigma_p^{})$ is attached to the dual cell $\check{\sigma}_{D-p-1}^{}$ to $\sigma_p^{}$.   
%
In particular, the gauge-invariance of $\hat{B}(\sigma_{p+1}^{})$ implies
\aln{
[\hat{A}(\sigma_{p-1}^{}),\hat{B}(\sigma_{p+1}^{})]=0\quad \text{for } ^\forall \sigma_{p-1}^{}\in \Delta_{p-1}^{}~\text{and } ^\forall \sigma_{p}^{}\in \Delta_{p}^{}~.
\label{commutativity}
}
%
 
Any gauge-invariant state $|\psi\rangle$ satisfies 
\aln{
\hat{G}(\sigma_{p-1}^{})|\psi\rangle=|\psi\rangle\quad \text{for } ^\forall \sigma_{p-1}^{}\in \Delta_{p-1}^{}~,
\label{gauge invariance}
}  
which means that $\hat{X}_{N}^{}(\sigma_{p-1}^{})$ and $\hat{A}(\sigma_{p-1}^{})$ are equivalent as operators acting on gauge-invariant states.       
In addition, we can fix the matter degrees of freedom as   
\aln{
\hat{Z}_N^{}(\sigma_{p-1}^{})|\psi\rangle=|\psi\rangle \quad \text{for all } \sigma_{p-1}^{}\in \Delta_{p-1}^{}~, 
}
by using the gauge degrees of freedom.  
As a result, the effective Hamiltonian becomes 
\aln{
\hat{H}=-g\sum_{\sigma_{p+1}^{}\in \Delta_{p+1}^{}}\hat{B}(\sigma_{p+1}^{})-\kappa \sum_{\sigma_p^{}\in \Delta_p^{}}\hat{Z}_N^{}(\sigma_p^{})-h\sum_{\sigma_p^{}\in \Delta_p^{}}\hat{X}_N^{}(\sigma_p^{})-g'\sum_{\sigma_{p-1}^{}\in \Delta_{p-1}^{}}\hat{A}(\sigma_{p-1}^{})+{\rm h.c.}~,
\label{effective Hamiltonian}
}
as an Hermitian operator acting on gauge-invariant states.   

In particular, the standard $\mathbb{Z}_N^{}$ $p$-form toric code corresponds to the limit $\kappa,h\rightarrow 0$,
\aln{
\hat{H}_{\rm TC}^{}\coloneq -g\sum_{\sigma_{p+1}^{}\in \Delta_{p+1}^{}}\hat{B}(\sigma_{p+1}^{})-g'\sum_{\sigma_{p-1}^{}\in \Delta_{p-1}^{}}\hat{A}(\sigma_{p-1}^{})+{\rm h.c.}~,
\label{toric code}
}
which consists of the $p$-dimensional Wilson surface operator~$\hat{B}^{}(\sigma_{p+1}^{})$ and the dual $(D-p-1)$-dimensional Wilson surface  operator $\hat{A}^{}(\sigma_{p-1}^{})$.  

\newpage

\noindent{\bf $p$ FORM GLOBAL SYMMETRY}
\begin{figure}
    \centering
     \includegraphics[scale=0.6]{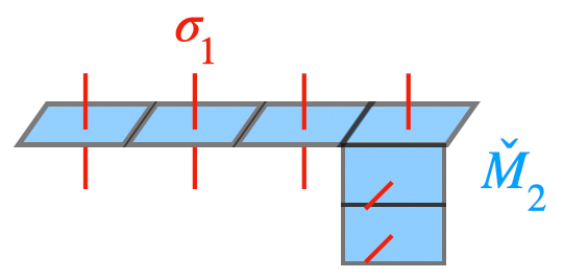}
    \caption{An example of a dual surface. 
    Here, we consider $D=3$ and $p=2$. 
    Red lines correspond to $1$-cells $\sigma_1^{}$ while blue surfaces correspond to dual $2$-cells.   
    }
    \label{fig:surface}
\end{figure}

\noindent The $\mathbb{Z}_N^{}$ $(p-1)$-form gauge symmetry in Eq.~(\ref{ZN gauge theory}) implies the existence of a $p$-form global symmetry $\mathbb{Z}_N^{[p]}$.  
The charged operator is a $p$-dimensional Wilson-surface operator 
\aln{
\hat{W}[C_p^{}]\coloneq \prod_{\sigma_{p+1}^{}\subset M_{p+1}^{}}\hat{B}^{}(\sigma_{p+1}^{})~,
}
where $C_p^{}$ is a $p$-dimensional closed subspace, $M_{p+1}^{}$ is an internal region of it, and $ \prod_{\sigma_{p+1}^{}\subset M_{p+1}^{}}$ is the product over all $(p+1)$-cells on $M_{p+1}^{}$.  
A symmetry operator can be constructed by the gauge-transformation operators as
\aln{
\label{symmetry operator}
\hat{U}_\omega^{}[\check{C}_{D-p-1}^{}]\coloneq \prod_{\substack{\sigma_{p-1}^{}~{\rm s.t.}\\
\mathrm{I}[\sigma_{p-1}^{},\check{M}_{D-p}^{}]\neq 0}} \hat{G}(\sigma_{p-1}^{})~,
}  
where $\check{C}_{D-p-1}^{}$ is a $(D-p-1)$-dimensional dual closed surface, $\check{M}_{D-p}^{}$ is the minimal surface bounded by $\check{C}_{D-p-1}^{}$, and $\mathrm{I}[\sigma_{p-1},\check{M}_{D-p}^{}]$ is their intersecting number in $\Delta_{D-1}^{}$. 
See Fig.~\ref{fig:surface} for example, where the case with $(D,p)=(4,2)$ is illustrated. 
In fact, one can check their commutation relation
\aln{
\hat{U}_\omega^{}[\check{C}_{D-p-1}^{}]^{}\hat{W}[C_p^{}]\hat{U}_\omega^{-1}[\check{C}_{D-p-1}^{}]=\omega^{-\mathrm{I}[C_p^{},\check{C}_{D-p-1}^{}]}\hat{W}[C_p^{}]~.
\label{commutation relation}
}
As we mentioned before, the intersection number is zero, $\mathrm{I}[C_p^{},\check{C}_{D-p-1}^{}]=0$, as long as we consider a flat spatial complex $\Delta_{D-1}^{}\sim \mathbb{R}^{D-1}$. 
On the other hand, if $\Delta_{D-1}^{}$ contains nontrivial subspaces along which $C_p^{}$ and/or $\check{C}_{D-p-1}^{}$ can wind, $\mathrm{I}[C_p^{},\check{C}_{D-p-1}^{}]$ can be nonzero, which then can lead to topological order. 
%
%
%

\subsection{Ground states}
In the following, we simply set $\kappa=0$, which corresponds to neglecting the matter sector.  
The fourth term on the right-hand side in Eq.~(\ref{effective Hamiltonian}) should then be interpreted as enforcing the gauge invariance in the gauge sector by taking $g'\rightarrow +\infty$.   
As a result, we arrive at the effective Hamiltonian  
\aln{
\hat{H}=-g\sum_{\sigma_{p+1}^{}\in \Delta_{p+1}^{}}\hat{B}(\sigma_{p+1}^{})
-h\sum_{\sigma_p^{}\in \Delta_p^{}}(\hat{X}_N^{}(\sigma_p^{})-1)+
{\rm h.c.}~,
\label{effective Hamiltonian 1}
} 
where we have replaced $\hat{X}_N^{}(\sigma_p^{})$ by $\hat{X}_N^{}(\sigma_p^{})-1$ so that the the energy of the trivial state $|0\rangle_X^{\otimes |\Delta_p^{}|}$ becomes zero. 

Let us now examine the ground state of this system.

\

\noindent {\bf UNBROKEN PHASE}\\
First, we focus on the parameter region $h\gg g\geq 0$. 
In this parameter region, the Hamiltonian is approximated by 
\aln{
\hat{H}\sim -h\sum_{\sigma_p^{}\in \Delta_p^{}}\hat{X}_N^{}(\sigma_p^{})+{\rm h.c.}~,
}
and the ground state is given by the trivial product state 
\aln{
|\text{unbroken}\rangle=|0\rangle_X^{\otimes |\Delta_p^{}|}~,
\label{unbroken state}
}
where $|\Delta_p^{}|$ is the total number of $p$-cells. 
The expectation value of the Wilson-surface operator $\langle \text{unbroken}|\hat{W}[C_p^{}]|\text{unbroken}\rangle$ is trivially zero, i.e., the $\mathbb{Z}_N^{}$ $p$-form global symmetry is unbroken. 
Note that the gauge invariance~(\ref{gauge invariance}) is trivially satisfied as well.

In this unbroken phase, an excited state can be obtained by acting $\hat{Z}_N^{}(\sigma_p^{})$ on a finite number of $p$-cells. 
However, the gauge invariance~(\ref{gauge invariance}) constrains such a configuration to form a $p$-dimensional closed intersecting surface $C_{p}^{}$  made by $(N-1)$ types of $p$-cells as  
\aln{
|C_p^{};\{k\}\rangle\coloneq 
\prod_{\sigma_{p+1}^{}\in \Delta_{p+1}^{}}(\hat{B}(\sigma_{p+1}^{}))^{m(\sigma_{p+1}^{})}|\text{unbroken}\rangle~,\quad m(\sigma_{p+1}^{})\in \{1,2,\cdots,N-1\}~,
\label{closed excitation}
}
which is gauge invariant due to the commutativity~(\ref{commutativity}).  
Here, $\{k\}=\{k(\sigma_p^{})\}_{\sigma_p^{}\in \Delta_p^{}}^{}$ is the set of  eigenvalues of $\frac{N}{2\pi i}\log \hat{X}_N^{}(\sigma_p^{})$ on each $p$-cell $\sigma_p^{}$:
\aln{
k(\sigma_p^{})=\sum_{\sigma_{p+1}^{}\in \Delta_{p+1}^{}}\langle \partial\sigma_{p+1}^{},\sigma_p^{}\rangle m(\sigma_{p+1}^{})^{}\quad \text{mod $N$}~
}
In particular, these eigenvalues satisfy 
\aln{
\sum_{\sigma_{p}\in \Delta_p^{}}\langle \sigma_p^{},\overline{\partial}\sigma_{p-1}^{}\rangle k(\sigma_{p}^{})\equiv 0 
\quad \text{mod $N$}~
} 
on each $\sigma_{p-1}^{}\in \Delta_{p-1}^{}$ by the nilpotency condition~(\ref{nilpotency2}).  
This is interpreted as the Gauss low in the present gauge theory.   
%
%
%
%
%
We should note that, even when the shape of $C_p^{}$ is fixed, a different assignment of $k(\sigma_p^{})$ on each $\sigma_{p}^{}$ leads to a different surface state $|C_p^{};\{k'\}\rangle$ for $N\geq 3$. 

\

\noindent {\bf BROKEN PHASE}\\
Next, let us focus on the parameter region $h\ll g$.  
This regime corresponds to the toric code~(\ref{toric code}) (with $g'\rightarrow \infty$). 
The gauge invariance~(\ref{gauge invariance}) implies that the ground state, denoted as $|\text{broken}\rangle$, is now a simultaneous eigenstate of both surface operators:
\aln{
\hat{A}(\sigma_{p-1}^{})|\text{broken}\rangle=|\text{broken}\rangle~,\quad 
\hat{B}(\sigma_{p+1}^{})|\text{broken}\rangle=|\text{broken}\rangle~. 
\label{ground state condition in toric code}
}  
The second condition indicates that the ground state cannot be a simple closed-surface state like Eq.~(\ref{closed excitation}), but is given by the equal-weight superposition of all $p$-dimensional closed surfaces as  
\aln{
|\text{broken}\rangle=\sqrt{{\cal N}}\sum_{\text{all}~C_p^{}}\sum_{\text{all}~\{k\}}|C_p^{};\{k\}\rangle~,\label{broken state}
}  
where $\sqrt{{\cal N}}$ is a normalization factor. 
Or by introducing the projection operator 
\aln{\hat{P}(\sigma_{p+1}^{})\coloneq \sum_{n=0}^{N-1}(\hat{B}(\sigma_{p+1}^{}))^n~,
} 
the above state can be also expressed as 
\aln{=\sqrt{\cal N}\prod_{\sigma_{p+1}^{}\in \Delta_{p+1}^{}}\hat{P}(\sigma_{p+1}^{})|0\rangle_X^{\otimes |\Delta_p^{}|}~.
}  
Obviously, this expression implies  
\aln{\langle\text{broken}| \hat{W}[C_p^{}]|\text{broken}\rangle=1~
}
for $^\forall C_p^{}\in \Gamma_p^{}$, indicating that $\mathbb{Z}_N^{[p]}$ is spontaneously broken. 
%

\subsection{Variational method}

Let us now apply the variational method developed in the previous sections to the $\mathbb{Z}_N^{}$ $p$-form gauge theory. 

The Wilson-surface operator $\hat{W}[C_p^{}]$ plays a role of the fundamental field $\hat{\phi}[C_p^{}]$ in the general framework, and the variational state is  
\aln{
|\psi_G^{}\rangle=\sqrt{{\cal N}}:\exp\left(\sum_{C_p^{}\in \Gamma_p^{}}~\hat{W}^\dagger [C_p^{}]\psi_G^{}[C_p^{}]
\right):|0\rangle_X^{\otimes |\Delta_p^{}|}~,
\label{variational state in gauge theory}
} 
where the prescription $:\cdots:$ is defined such that each surface state appears with equal weight when expanding the exponential.   
%
For example, let us focus on a surface state which takes a form of  
\aln{
\cdots \hat{W}[C_p^{}]\hat{W}[C_p^{'}]|0\rangle_X^{\otimes |\Delta_p^{}|}~,\quad C_p^{}~,~C_p^{'}\in \Gamma_p^{}~,
}
where $\cdots$ denotes other surface contributions. 
When $C_p^{}+C_p^{'}\in \Gamma_p^{}$, however, this is identical to 
\aln{
\cdots \hat{W}[C_p^{}+C_p^{'}]|0\rangle_X^{\otimes |\Delta_p^{}|}~ 
} 
by the fusion property of the Wilson-surface operator $\hat{W}[C_p^{}]\hat{W}[C_p^{'}]=\hat{W}[C_p^{}+C_p^{'}]$. 
In this way, a naive expansion of the exponential in Eq.~(\ref{variational state in gauge theory}) contain infinitely many identical states, and the prescription $:\cdots:$ is therefore introduced to count these states equally.  
%
By definition, we have  
\aln{
\hat{W}[C_p^{}]|\psi_G^{}\rangle=\psi_G^{}[C_p^{}]|\psi_G^{}\rangle~
\label{coherent property}
} 
for $^\forall C_p^{}\in \Gamma_p^{}$.\footnote{
In this case, this coherence property originates from the concatenation property of the Wilson surface $W[C_p^{}](W^\dagger[C_p^{}])^n=(W^\dagger[C_p^{}])^{n-1}$ rather than the commutation relation.  
} 
%
%
Furthermore, this coherence property leads to additional constraint 
\aln{\psi_G^{}[C_p^{}]^n=1
\label{condition for wave function}
}
for $^\forall C_p^{}\in \Gamma_p^{}$ if $\psi_G^{}[C_p^{}]\neq 0$.  
Accordingly, the variation of free-energy functional must be taken under this condition.

In order to examine the expectation value of the Hamiltonian~(\ref{effective Hamiltonian 1}) in the above variational state, we first need to deform the kinetic term in an appropriate way.   
For this purpose, we represent the infinitesimal surface deformation~(\ref{functional variation}) in an operator form as ~\cite{Yoneya:1980bw,Banks:1980sq} 
\aln{
\Pi_{\pm \partial \sigma_{p+1}^{}}^{}\hat{W}[C_p^{}]\coloneq \begin{cases}\hat{W}[C_p^{}\pm \partial \sigma_{p+1}^{}] & \text{for } C_p^{}\pm \partial\sigma_{p+1}^{}\in \Gamma_p^{}
\\
0 & \text{for } C_p^{}\pm \partial\sigma_{p+1}^{}\notin \Gamma_p^{}
\end{cases}~.
\label{translation operator}
}
Using this operator, we can express the kinetic term in Eq.~(\ref{effective Hamiltonian 1}) as
\aln{
-g \sum_{\sigma_{p+1}\in \Delta_{p+1}^{}}(\hat{B}(\sigma_{p+1}^{})+\hat{B}^\dagger(\sigma_{p+1}^{}))&=-g\sum_{C_{p}^{}\in \Gamma_p^{}}w[C_p^{}]\hat{W}^\dagger[C_p^{}]{\cal H}\hat{W}[C_p^{}]
~,
\label{quadratic kinetic term}
}
where the operator ${\cal H}$ is defined by 
\aln{
{\cal H}\hat{W}[C_p^{}]\coloneqq 
\sum_{\substack{\sigma_{p+1}^{}\in \Gamma_p^{}}}\sum_{s=\pm}\Pi_{s\partial\sigma_{p+1}^{}}^{}\hat{W}[C_p^{}]~.
\label{derivative operator}
}
This is proved as follows: 
\aln{
\sum_{C_p^{}\in \Gamma_p^{}} w[C_p^{}]\hat{W}^\dagger [C_p^{}]{\cal H}w[C_p^{}] \hat{W}[C_p^{}]
&=\sum_{C_p^{}\in \Gamma_p^{}}w[C_p^{}]\hat{W}^\dagger [C_p^{}]
\sum_{s=\pm }\sum_{\substack{\sigma_{p+1}^{}~{\rm s.t.}\\ C_p^{}+s\partial \sigma_{p+1}^{}\in \Gamma_p^{}}}
\hat{W}[C_p^{}+s\partial \sigma_{p+1}^{}]
\nn
&=\sum_{\sigma_{p+1}^{}\in \Delta_{p+1}^{}}\sum_{s=\pm}\hat{B}(s\sigma_{p+1}^{})\sum_{\substack{C_p^{}~{\rm s.t.}\\ C_p^{}+s \partial \sigma_{p+1}^{}\in \Gamma_p^{}}}
w[C_p^{}]
\nn
&=\sum_{\sigma_{p+1}^{}\in \Delta_{p+1}^{}}(\hat{B}(\sigma_{p+1}^{})+\hat{B}^\dagger(\sigma_{p+1}^{}))~,
\label{proof}
}
where we have used Eq.~(\ref{concrete w condition}) from the second line to the third line. 
%
%
Then, using Eq.~(\ref{coherent property}), the expectation value of the kinetic term can be evaluated as  
\aln{
&-g\left\langle \psi_G^{}\bigg|\sum_{\sigma_{p+1}^{}\in \Delta_{p+1}^{}}\left(\hat{B}(\sigma_{p+1}^{})+\hat{B}^\dagger (\sigma_{p+1}^{})\right)\bigg |\psi_G^{}\right\rangle
=-g\sum_{C_p^{}\in \Gamma_p^{}}w[C_p^{}]\psi_G^{}[C_p^{}]^*{\cal H}\psi_G^{}[C_p^{}]~.
\label{naive kinetic term}
}
Similarly, the expectation value of the second term in Eq.~(\ref{effective Hamiltonian 1}) is 
\aln{
\left\langle \psi_G^{}\bigg| \sum_{\sigma_p^{}\in \Delta_p^{}}\left(\hat{X}_N^{}(\sigma_p^{})-1\right)\bigg|\psi_G^{}\right\rangle+{\rm h.c.}
&=\left\langle \psi_G^{}\bigg|\sum_{C_p^{}\in \Gamma_p^{}}\sum_{\sigma_p^{}\in C_p^{}}\left(\omega^{-\langle \partial\sigma_{p+1}^{},\sigma_p^{}\rangle}-1\right)\hat{W}^\dagger [C_p^{}]\psi_G^{}[C_p^{}]\bigg|\psi_G^{}\right\rangle+{\rm h.c.}
\nn
&=-4\left(\sin\frac{\pi}{N}\right)^2\sum_{C_p^{}\in \Gamma_p^{}}|C_p^{}|\psi_G^{}[C_p^{}]^*\psi_G^{}[C_p^{}]~,
} 
where we have used  
\aln{
\left(\sum_{\sigma \in \Delta_p^{}}(\hat{X}_N^{}(\sigma_p^{})-1)\right)\hat{W}^\dagger[C_p^{}]|0\rangle_X^{\otimes |\Delta_p^{}|}=\sum_{\sigma_p^{}\in C_p^{}}\left(\omega^{-\langle \partial\sigma_{p+1}^{},\sigma_p^{}\rangle}-1\right)\hat{W}^\dagger[C_p^{}]|0\rangle_X^{\otimes |\Delta_p^{}|}
}
for $^\forall C_p^{}\in \Gamma_p^{}$. 
In the following, the overall coefficient $4\left(\sin\frac{\pi}{N}\right)^2$ is absorbed into $h$. 

Collecting all terms and taking the constraint~(\ref{condition for wave function}) into account, the free energy functional is given by\footnote{
Here, we pretend that $w[C_p^{}]$ is a constant, namely, given by Eq.~(\ref{simplest weight}) so that it can be factored out as an overall  factor.
}
\aln{
F[\psi_G^{}]&=\langle \psi_G^{}|\hat{H}|\psi_G^{}\rangle+\lambda \sum_{C_p^{}\in \Gamma_p^{}}w[C_p^{}]\left(\psi_G^{}[C_p^{}]^N-1\right)+{\rm h.c.} 
\nn
&=\sum_{C_p^{}\in \Gamma_p^{}}w[C_p^{}]\bigg\{-g\psi_G^{}[C_p^{}]^*\left({\cal H}-n[C_p^{}]\right)\psi_G^{}[C_p^{}]
+|C_p^{}|\left(h-g\frac{n[C_p^{}]}{|C_p^{}|}\right)\psi_G^{}[C_p^{}]^*\psi_G^{}[C_p^{}]
\nn
&\hspace{9cm}+\lambda\psi_G^{}[C_p^{}]^N+{\rm h.c.}
\bigg\}~,
\label{energy functional in ZN theory}
}
where $n[C_p^{}]~(\propto |C_p^{}|)$ is the total number of $(p+1)$-cells which share at least one $p$-cell with $C_p^{}$.\footnote{For example, when $\Delta_{D-1}^{}$ is a hyper-cubic lattice, one finds $n[C_p^{}]=2(D-p-1)|C_p^{}|$, where $(D-p-1)$ is the codimension of $C_p^{}$ in $\Delta_{D-1}^{}$ and the factor of $2$ accounts for the choice of orientations.  
These factors depend on $\Delta_{D-1}^{}$, whereas the volume dependence $n[C_p^{}]\propto |C_p^{}|$ is generic. 
}  
Here, readers might question why $n[C_p^{}]$ is subtracted from the kinetic term; we will see that this subtraction leads to a correct form of the kinetic term in the continuum limit. 
%

Now, we can perform a mean-field analysis as before.   
\

\noindent {\bf GROUND STATES}\\ 
First, let us see how the ground states obtained in the previous subsection are reproduced in the present variational approach.  
As long as we focus on ground states, we can neglect the kinetic term in Eq.~(\ref{energy functional in ZN theory}). 
In the weak gauge-coupling regime $h\gg g\geq 0$, the effective potential is 
\aln{|C_p^{}|h\psi_G^{*}\psi_G^{}+\lambda \psi_G^{N}+{\rm h.c}~,\quad h,\lambda>0~,
}
whose global minimum exists at $\psi_G^{}=0$. 
Correspondingly, the ground state is 
\aln{
|\psi_G^{}\rangle=|0\rangle_X^{\otimes |\Delta_p^{}|}~, 
} 
which coincides with $|\text{unbroken}\rangle$ given in Eq.~(\ref{unbroken state}). 

On the other hand, in the strong gauge-coupling limit $g\gg h\geq 0$, 
the effective potential is 
\aln{-ag\psi_G^{*}\psi_G^{}+\lambda \psi_G^{N}+{\rm h.c}~,
}
where $a=n[C_p^{}]/|C_p^{}|$ is a positive constant. 
This has $N$ degenerate minima, 
\aln{\psi_G^{}=v\times \omega^m~,\quad v\neq 0~,\quad m=0,1,\cdots,N-1~,
}
among which $m=0$ corresponds to the ground state 
\aln{
|\psi_G^{}\rangle^{}=\sqrt{\cal N}:\exp\left(v\sum_{C_p^{}\in \Gamma_p^{}}\hat{W}^\dagger [C_p^{}]\right):|0\rangle_X^{\otimes |\Delta_p^{}|}~,
\label{N ground states}
}
which coincides with $|\text{broken}\rangle$ as given in Eq.~(\ref{broken state}). 
Thus, this represents a condensation of surfaces and corresponds to the   spontaneously broken phase of $\mathbb{Z}_N^{[p]}$.  

\

\noindent {\bf SURFACE EXCITATIONS}\\
Surface excitations can be also constructed by following the general argument in Section~\ref{discrete symmetry}. 
In the present $\mathbb{Z}_{N}^{}$ gauge theory, the open surface operator $\hat{W}[M_p^{}]~(\hat{U}_\omega^{}[\check{M}_{D-p-1}^{}])$ is obtained from $\hat{W}[C_p^{}]~(\hat{U}_\omega^{}[\check{C}_{D-p-1}^{}])$ by removing $\hat{Z}_N^{\pm}(\sigma_p^{})~(\hat{X}_N^{\pm}(\sigma_p^{}))$ on an arbitrary subset of $\sigma_p^{}\in C_p^{}~(\check{\sigma}_{D-p-1}^{}\in \check{C}_{D-p-1}^{})$. 
Explicitly, they are given by
\aln{
\hat{W}[M_{p}^{}]&=\hat{W}[C_p^{}]\prod_{\sigma_p^{}\in C_p^{}\backslash M_p^{}}\hat{Z}_N^{-\langle \partial\sigma_{p+1}^{},\sigma_p^{}\rangle}(\sigma_p^{})~,\quad \partial M_p^{}\equiv C_{p-1}^{}\neq \emptyset~,
\\
\hat{U}_\omega^{}[\check{M}_{D-p-1}^{}]&=\hat{U}_\omega^{}[\check{C}_{D-p-1}^{}]\prod_{\check{\sigma}_{D-p-1}^{}\in \check{C}_{D-p-1}^{}\backslash \check{M}_{D-p-1}^{}}\hat{X}_N^{-\langle \partial\sigma_{p+1}^{},\sigma_p^{}\rangle}(\sigma_p^{})~,\quad \partial \check{M}_{D-p-1}^{}\equiv \check{C}_{D-p-2}^{}\neq \emptyset~,
}  
where $C_{p-1}^{}~(\check{C}_{D-p-2}^{})$ is a (dual) $(p-1)$-($(D-p-2)$-)dimensional closed subspace in $\Delta_{D-1}^{}~(\check{\Delta}_{D-1}^{})$.
As discussed before, the states
\aln{
|e\rangle_{p-1}^{}\coloneq \hat{W}[M_{p}^{}]|\psi_G^{}\rangle^{}~,\quad |m\rangle_{D-p-2}^{}\coloneq \hat{U}_w^{}[\check{M}_{D-p-1}^{}]|\psi_G^{}\rangle^{}~,
} 
represent boundary excitations, and acquire fractional braiding phases when they are linked each other. 
Note that the origin of such braiding phases lies in the commutation relation between $\hat{Z}_N^{}(\sigma_p^{})$ and $\hat{X}_N^{}(\sigma_p^{})$ in contrast to the commutation relation of $\hat{\phi}[C_p^{}]$ and $\hat{\phi}^\dagger[C_p^{}]$ in Section~\ref{mean field analysis}. 

\

\noindent {\bf CONTINUUM LIMIT}\\
Finally, let us discuss the continuum limit of the variational free energy~(\ref{energy functional in ZN theory}). 
Since low-energy effective theory would be independent of microscopic details, we here choose a hypercubic lattice with a lattice spacing $a$ as $\Delta_{D-1}^{}$.     
Then, employing the projected area derivative~(\ref{projected area derivative}), the kinetic term in Eq.~(\ref{energy functional in ZN theory}) can be expanded as
%
\aln{
&\psi_G^{}[C_p^{}]^*\left({\cal H}-n[C_p^{}]\right)\psi_G^{}[C_p^{}]
\nn
=&\sum_{\hat{i}\in C_p^{}}\psi_G^{}[C_p^{}]^*\left(a^{p+1}\sum_{s=\pm}s\sum_{k=1}^{D-1}\frac{\delta \psi_G^{}[C_p^{}]}{\delta \Sigma^{k}(\hat{i})}
+\frac{(a^{p+1})^2}{2}\sum_{s=\pm}s^2\sum_{k=1}^{D-1}\frac{\delta^2\psi_G^{}[C_p^{}]}{\delta \Sigma^{k}(\hat{i})\delta \Sigma_{k}^{}(\hat{i})}+{\cal O}((a^{p+1})^3)
\right)
\nn
=&(a^{p+1})^2\sum_{\hat{i}\in C_p^{}}\psi_G^{}[C_p^{}]^*\sum_{k=1}^{D-1}\frac{\delta^2\psi_G^{}[C_p^{}]}{\delta \Sigma^{k}(\hat{i})\delta \Sigma_{k}^{}(\hat{i})}+{\cal O}((a^{p+1})^3)~,
}
where the linear term in the projected area derivatives vanishes due to two  contributions with opposite orientations, $s=\pm 1$.       
Then, the continuum limit of Eq.~(\ref{energy functional in ZN theory}) is  
\aln{
F[\psi_G^{}]=\int [dC_p^{}]&\left\{-g\psi_G^{}[C_p^{}]^*\Box_p^{}\psi_G^{}[C_p^{}]-\mu \psi_G^{}[C_p^{}]^*\psi_G^{}[C_p^{}]+\frac{\lambda}{\mathrm{Vol}[C_p^{}]}\psi_G^{}[C_p^{}]^N+{\rm h.c.}
\right\}
\nn
&+(\text{higher derivative terms})~,
}
which is identical to Eq.~(\ref{continuum surface energy}).  
Thus, we can repeat the same analysis in Section~\ref{discrete symmetry} and find  that low-energy effective theory is described by the $\mathrm{BF}$-type topological field theory, which can couple to topological defects as in Eq.~(\ref{effective theory in discrete}). 
%

\section{Summary and discussion}\label{sec:summary}

We have investigated the variational method for interacting surface systems with higher-form global symmetries. 
Analogously to conventional bosonic systems, we constructed a second-quantized Hamiltonian for the surface operator $\hat{\phi}[C_p^{}]$, which possesses a higher-form global symmetry $G^{[p]}$.
 Taking the expectation value of the Hamiltonian in a variational coherent state, we obtained the free-energy functional, whose variation leads to the functional Schr\"{o}dinger equation, analogous to the Gross-Pitaevskii equation in conventional bosonic systems.
Based on this generalized Gross-Pitaevskii equation, we have performed the mean-field analysis and uncovered various physical consequences associated with the spontaneous breaking of higher-form global symmetries. 
Finally, as a concrete microscopic model, we studied a $\mathbb{Z}_N^{}$ $p$-form lattice gauge theory and applied our variational framework in detail.    

Many important issues remain to be addressed. 
First, one can also consider gauging our second-quantized system by introducing a $(p+1)$-form gauge field $A_{p+1}^{}$. 
Such a theory correspond to a higher-form generalization of the conventional Ginzburg-Landau theory and is expected to describe a superconducting phase of extended objects~\cite{Kawana:2024fsn}. 

Second, as a related problem, it seems possible to extend our formulation to fermionic surface systems by promoting the fundamental field $\hat{\phi}[C_p^{}]$ to a fermionic surface operator that obeys the canonical anti-commutation relation, $\{\phi[C_p^{}],\hat{\phi}^\dagger [C_p^{'}]\}=w[C_p^{}]^{-1}\delta_{C_p^{},C_p^{'}}^{}$.  
It is particularly interesting to ask whether we can generalize the BCS theory for interacting fermions to such a fermionic theory of surfaces with higher-form global symmetries.    
In addition, it would be of great interest to construct and analyze    supersymmetric theories by incorporating both bosonic and fermionic surface operators.  

Third, applying our variational method to systems with mobility constrains is also a fruitful research direction~\cite{PhysRevLett.94.040402,Haah:2011drr,Vijay:2016phm,Hirono:2022dci,Ebisu:2023idd,Ebisu:2024cke}. 
In such systems, higher-form global charges are not commutative with certain spacial translation operators, leading to the emergence of fracton excitations in the low-energy limit. 
It is quite interesting to investigate whether effective field theories of these fractons can be obtained by applying our variational approach as presented in this paper.        

We hope to discuss these problems in the near future.

\section*{Acknowledgements}
%
This work is supported by KIAS Individual Grants, Grant No. 090901.   
%


\bibliographystyle{TitleAndArxiv}
\bibliography{Bibliography}

\end{document}